# Kinetic Solvers with Adaptive Mesh in Phase Space

Robert R. Arslanbekov,[a] Vladimir I. Kolobov,[a] and Anna A. Frolova[b]

[a]*CFD Research Corporation, 215 Wynn Dr, Huntsville, AL 35803, USA*
[b]*Dorodnitsyn Computing Center of the Russian Academy of Sciences,
Vavilova Str., 40, Moscow, 119333, Russia*

**Abstract.** An Adaptive Mesh in Phase Space (AMPS) methodology has been developed for solving multi-dimensional kinetic equations by the discrete velocity method. A Cartesian mesh for both configuration (r) and velocity (v) spaces is produced using a "Tree of Trees" (ToT) data structure. The r-mesh is automatically generated around embedded boundaries, and is dynamically adapted to local solution properties. The v-mesh is created on-the-fly for each r-cell. Mappings between neighboring v-space trees is implemented for the advection operator in r-space. We have developed new algorithms for solving the full Boltzmann and linear Boltzmann equations with AMPS. Several recent innovations were used to calculate the discrete Boltzmann collision integral with dynamically adaptive v-mesh: the importance sampling, multi-point projection, and variance reduction methods. We have developed an efficient algorithm for calculating the linear Boltzmann collision integral for elastic and inelastic collisions of hot light particles in a Lorentz gas. The new AMPS technique has been demonstrated for simulations of hypersonic rarefied gas flows, ion and electron kinetics in weakly ionized plasma, radiation and light particle transport through thin films, and electron streaming in semiconductors. We have shown that AMPS allows minimizing the number of cells in phase space to reduce the computational cost and memory usage for solving challenging kinetic problems.

**Keywords:** Tree-of-Trees, Boltzmann kinetic equation, Rarefied Gas Dynamics, Lorentz gas, Discrete Velocity Method, Adaptive Mesh in Phase Space, Hypersonic Flows, Electron kinetics in gas discharges and semiconductors.

## I. INTRODUCTION

Kinetic equations are widely used in many fields from physics to sociology and finances [1]. Typical applications in physical kinetics include rarefied gas dynamics, radiation and heat transport, and the kinetics of charged particles in plasmas and semiconductors [2,3]. Two major methods for solving kinetic equations are statistical particle-based simulation methods and direct numerical solutions using computational mesh in phase space, which includes physical and velocity spaces. In the latter case, a Discrete Velocity Method (DVM) is used for discretizing velocity space [4,5,6] to resolve VDF shape for all points of physical space. The typical number of velocity cells in DVM is much larger than the number of discrete velocities used in the Broadwell models [7] and Lattice Boltzmann Methods (LBM) [8], both of which operate with a minimal number of discrete velocities to simulate dynamics close to equilibrium.

Adaptive Mesh Refinement (AMR) has been used for solving Partial Differential Equations (PDE) with reduced number of computational cells [9]. In particular, adaptive Cartesian meshes based on hierarchical data structures [10] have gained popularity in computational science [11]. The benefits of AMR increase sharply with increasing dimensionality of the problem, which makes AMR highly desirable for multi-dimensional kinetic solvers. Attempts to use Adaptive Mesh in Phase Space (AMPS) for the kinetic equations can be found in recent literature [12,13,14,15]. AMR allows resolving important regions of phase space where the particles are present and reduce the number of cells in the regions with no particles. This makes DVM resembling the particle-based methods which need no mesh for particle tracing. Extending to velocity space the AMR techniques developed for the physical (configuration) space could provide substantial savings in computational cost. In particular, AMPS could drastically increase the efficiency of direct kinetic solvers for problems with large variations of the velocity distribution functions (VDF) in phase space.

Similar ideas have been evolving in the LBM community. The LBM, originally designed as an alternative solver for computational fluid dynamics (CFD), has been extended beyond the level of the Navier Stokes hydrodynamics and capable of describing some kinetic effects [16,17]. It is a mesoscopic method, which utilizes discrete values of the VDF on a minimal set of discrete velocities to obtain governing equations for fluid dynamics alternative to conservation equations based on VDF moments. Most LBM works are devoted to low speed isothermal flows close to equilibrium. However, LBM with a larger number of discrete velocities [18,19], adaptive meshes in physical space [20], and finite volume (FV) LBM with unstructured meshes [21,17] have been recently developed to expand LBM capabilities. The ideas of using locally adaptive velocity sets for the Broadwell models and LBM have been described in the literature [22]. In particular, decomposing particle velocity into a (locally adaptive) mean flow velocity and a so-called peculiar velocity is one of the methods enabling LBM extensions for compressible flows [23]. In several aspects these methods resemble the computational technologies described in the present paper.

We have previously developed a Unified Flow Solver (UFS) for simulations of gas flows over a wide range of Knudsen and Mach numbers [24]. UFS uses Adaptive Mesh and Algorithm Refinement (AMAR) methodologies [25], which combine AMR with dynamic selection of kinetic and fluid solvers in different parts of computational domain based on continuum breakdown criteria. The Boltzmann equation is solved by splitting free flight and collisions with explicit, first-order accurate time marching scheme. The original Boltzmann solver in UFS uses a structured Cartesian grid in velocity space. This grid is static (does not change with time) and is the same for all cells in configuration space (global velocity mesh). For highly non equilibrium gas flow problems at large Mach and large Knudsen numbers and large variations of gas temperatures, the VDF varies drastically in phase space and possibly in time. This is typical to shock waves and boundary layers in hypersonic rarefied gas dynamics as well as high speed microflows. Solutions of such problems require velocity grids which are large in size (to cover the entire range of possible velocities) and dense (to resolve smallest VDF features). Using uniform, static velocity grids leads to prohibitive memory and CPU requirements for these problems in multi-dimensional cases.

Adaptive Cartesian mesh in velocity space has been previously demonstrated for spatially homogeneous kinetic equations with different types of collision integrals [26,27]. Block-structured AMR algorithms for solving the Vlasov equation in 1D1V have been described in [13], where further references to previous works on Vlasov solvers can be found. A recent paper [14] describes the solution of the BGK kinetic equation in 2D2V using a moving mesh in configuration space and quadtree Cartesian mesh in velocity space. The first demonstration of adaptive Cartesian mesh in phase space for hypersonic rarefied flows has been recently presented using the BGK model for 2D2V settings [28]. The need for AMR techniques in both configuration and velocity spaces was expressed in a recent paper [29] devoted to large-scale space weather simulations with a six-dimensional Vlasov solver on multi-GPU clusters.

In the present paper, we introduce a new concept of a Tree-of-Trees (ToT) for solving multi-dimensional kinetic equations with adaptive phase space mesh. In this technique, tree-based adaptive Cartesian meshes are generated for both configuration and velocity spaces. The mesh in configuration space is refined around embedded objects of complex shape and dynamically adapted to local solution properties. A quad/octree mesh in velocity space is created on-the-fly for each cell in configuration space. The kinetic equations are solved by splitting the configuration and velocity grids and using an explicit time marching scheme. Kinetic solvers without splitting the space-velocity grids for phase spaces of small dimensions (up to three) are compared with those using the ToT technique (split grids). Problems associated with consistent mesh adaptation in configuration and velocity space are discussed. The benefits of the AMPS technique are demonstrated for hypersonic rarefied flows, radiation and light-particle transport in a Lorentz gas, and charged particle kinetics in plasmas and semiconductors.

The structure of the present paper is as follows. Section II provides an introduction to the kinetic equations and description of phase space required for understanding of the proposed AMPS method. We describe key differences between the DVM and LBM methods, which are mostly related to selection of discrete velocity sets. Section III introduces a general Tree-of-Trees framework developed for solution of the kinetic equations. We describe the VDF reconstruction (mapping) technique required for locally adaptive discrete velocity grids. In Section IV, we describe implementation of discrete collision integrals on adaptive velocity grids focusing on bi-linear Boltzmann collision integral for rarefied gas dynamics, and linear Boltzmann-Lorentz collision integrals for light species in a Lorentz gas. Section V demonstrates examples of simulations with the newly developed technique. We consider two types of problems. The first type includes hypersonic rarefied gas flows. The second type deals with kinetics of light particles in a binary mixture of gases with disparate mass of species (the Lorentz gas). We also use the Vlasov equation to illustrate differences between the split and unsplit space-velocity grid techniques. Conclusions are drawn in Section VI.

## II. KINETIC EQUATIONS AND PHASE SPACE

### A. Kinetic Equations

Kinetic equations of interest can be written in a conservative form suitable for FV discretization:

$$\frac{\partial f}{\partial t} + \nabla \cdot (\mathbf{A}f) = I, \qquad (1)$$

where $\nabla$ denotes the divergence of the particle flux in six-dimensional phase space $(\mathbf{r}, \boldsymbol{\xi})$, and $f(\mathbf{r}, \boldsymbol{\xi}, t)$ is the velocity distribution function (VDF), which depends on the position vector $\mathbf{r}$ in configuration space, and velocity vector $\boldsymbol{\xi}$. Vector $\mathbf{A}$ has two components $(\boldsymbol{\xi}, \mathbf{a})$, where $\mathbf{a}$ is the acceleration vector due to external forces. It is assumed that $\mathbf{a}$ does not depend on $\boldsymbol{\xi}$, so that $\nabla_\xi \cdot (\mathbf{a} f) = \mathbf{a} \cdot \nabla_\xi f$.[1] The left hand side of Eq. (1) describes a six-dimensional advection of an incompressible fluid with the phase space density, $f(\mathbf{r}, \boldsymbol{\xi}, t)$. For binary collisions, the collision term, $I$, is a local operator in configuration space. Below, we focus on the Boltzmann, Vlasov, and the Lorentz-Boltzmann equations.

The Boltzmann collision integral for elastic collisions of two atoms in a one-component gas has the following form [30]:

$$I = \int_{R^3} d\boldsymbol{\xi_1} \int_{S^2} (f(\boldsymbol{\xi_1'}) f(\boldsymbol{\xi'}) - f(\boldsymbol{\xi_1}) f(\boldsymbol{\xi})) |g| \sigma(|g|, \theta) d\Omega. \qquad (2)$$

Here, $\mathbf{g} = \boldsymbol{\xi_1} - \boldsymbol{\xi}$ is the vector of relative velocity of the colliding particles, and $d\Omega = \sin\theta d\theta d\phi$. The differential collision cross section, $\sigma(g, \theta)$, depends on the scattering angle between the relative particle velocities before and after collision. The particle velocities after collision, $(\boldsymbol{\xi'}, \boldsymbol{\xi_1'})$, and those prior to collision, $(\boldsymbol{\xi}, \boldsymbol{\xi_1})$, satisfy the laws of conservation of momentum and energy:

$$\begin{aligned}\boldsymbol{\xi} + \boldsymbol{\xi_1} &= \boldsymbol{\xi'} + \boldsymbol{\xi_1'} \\ |\boldsymbol{\xi}|^2 + |\boldsymbol{\xi_1}|^2 &= |\boldsymbol{\xi'}|^2 + |\boldsymbol{\xi_1'}|^2\end{aligned}. \qquad (3)$$

Elastic collisions redistribute velocity vectors on a sphere with the center $\boldsymbol{\xi_0} = (\boldsymbol{\xi} + \boldsymbol{\xi_1})/2$ and radius $g/2$ in velocity space. To determine the post-collision velocities $(\boldsymbol{\xi'}, \boldsymbol{\xi_1'})$, it is necessary to know the scattering angle which in turn depends on the interaction potential of the atoms. For a hard sphere (HS) model used in this paper, scattering is isotropic, $\sigma(g, \theta) = d^2/4$, where $d$ is the atomic diameter. For a soft sphere model used for simulations of granular flows only the first equation (moment conservation) in (3) is satisfied and energy dissipation takes place [3].

The lowest velocity moments of the VDF are the macroscopic variables, the number density, $N$, mean velocity, $\mathbf{v}$, temperature, $T$, pressure tensor, $\mathbf{p}$, and heat flux, $\mathbf{q}$:

$$N = \int f d\boldsymbol{\xi}, \qquad \mathbf{v} = \frac{1}{N} \int \boldsymbol{\xi} f d\boldsymbol{\xi}, \qquad T = \frac{M}{3 k_B N} \int c^2 f d\boldsymbol{\xi},$$

$$p_{ij} = M \int c_i c_j f d\boldsymbol{\xi}, \qquad \mathbf{q} = \frac{M}{2} \int c^2 \mathbf{c} f d\boldsymbol{\xi}.$$

Here $\mathbf{c} = \boldsymbol{\xi} - \mathbf{v}$, $M$ is the molecular mass, and $k_B$ is the Boltzmann constant. The subscripts $i$ and $j$ denote the Cartesian components. The density, flow velocity and temperature satisfy the conservation equations of mass, momentum, energy, etc., which are expressed by the Euler, Navier-Stokes, Burnet, and the Grad 13-moment equations. Higher moments, such as the fourth moments, $p_{2ii} = \int c_i^4 f(\boldsymbol{\xi})d\boldsymbol{\xi}$, are often used to illustrate accuracy of computational methods and the degree of system deviation from an equilibrium state.

---

[1] Since $\mathbf{r}$ and $\boldsymbol{\xi}$ are independent variables, the commutativity $\nabla_r \cdot (\boldsymbol{\xi} f) = \boldsymbol{\xi} \cdot \nabla_r f$ is always satisfied.

For gas mixtures, kinetic equations can be written for each component. Below, we consider a binary mixture of heavy and light species, which is called a Lorentz gas. The density and mass of the light species are denoted as *n* and *m*, respectively. For collisions between light and heavy particles in the Lorentz gas, the collision operator is linear, and can be obtained for a prescribed distribution of heavy (or target) species. For instance, elastic collisions of hot electrons with cold atoms, the leading term of the collision operator (at $m/M \ll 1$) has the Boltzmann-Lorentz form [31]:

$$I = N\xi \int_{S^2} \sigma(\xi,\theta)\left[f(\xi\boldsymbol{\Omega}') - f(\xi\boldsymbol{\Omega})\right]d\boldsymbol{\Omega}', \tag{4}$$

where the velocity vector $\boldsymbol{\xi} = \xi\boldsymbol{\Omega}$ is decomposed into its modulus, $\xi$, and the angular part, $\boldsymbol{\Omega}$. The differential collision cross section, $\sigma(\xi,\theta)$, is a function of the particle speed and the scattering angle $\theta = \arccos(\boldsymbol{\Omega}\cdot\boldsymbol{\Omega}')$ between the initial and final particle velocity during collision. The collision operator (4) modifies a direction of the particle's velocity (momentum) but conserves its modulus or kinetic energy.

Inelastic collisions are those in which hot light particles (e.g., electrons) lose a large fraction of their energy. The inelastic collisions of electrons with atoms or molecules are described by the collision operator [32]:

$$I = N\xi \sum_k \int_{S^2} \left[f(\zeta)\frac{\zeta^2}{\xi^2}\sigma_k(\zeta,\theta) - f(\xi)\sigma_k(\xi,\theta)\right]d\boldsymbol{\Omega}'. \tag{5}$$

Here $\boldsymbol{\zeta}$ is the post-collision velocity, $\zeta^2 = \frac{2\varepsilon_k}{m} + \xi^2$, where $\varepsilon_k$ is the energy required to excite a *k*-th state of an atom or molecule, $\sigma_k(\xi,\theta)$ is the differential cross-section for excitation of a *k*-th energy state, and $\sum_k$ denotes summation over all excited states.

## B. Discretization of Velocity and Configuration Spaces

For the numerical solution of the kinetic equations, we first discretize velocity space to obtain a discrete Boltzmann equation (in the absence of external forces) [7]:

$$\frac{\partial f_i}{\partial t} + \boldsymbol{\xi}_i \cdot \nabla_r f_i = \frac{1}{2}\sum_{j,k,l} A_{ij}^{kl}\left(f_k f_l - f_i f_j\right). \tag{6}$$

Here, $f_i$ denotes the value of VDF at the discrete velocity $\boldsymbol{\xi}_i$, and $A_{ij}^{lk}$ is a scattering matrix. For practical applications of Eq. (6), there are two major challenges: a) the discrete collision operator involves a summation over all discrete velocities, which is expensive (of the order of $N_v^2$, where $N_v$ is the number of discrete velocities), b) the discrete collision operator takes into account only those discrete post-collision velocities, which fall exactly on the collision sphere. The number of such velocities is small and strongly depends on the radius of the collisional sphere. To overcome these challenges (which become even more important for locally adaptive discrete velocity sets), several methods have been described in the literature (see Section IV).

Selection of discrete velocity sets is different for DVM and LBM [33]. In DVM, discrete velocities are selected to resolve the VDF shape. For general 3V cases, the typical number of cells in velocity space, $N_v$, is of the order of $10^3$. In LBM, the discrete values of the VDF are used as the state variables instead of the moment integrals to describe gas dynamics near equilibrium. The minimal set of discrete velocities is selected to satisfy mass, momentum and energy conservation, as well as rotational symmetry. In particular, LBM models are dubbed DnQm for *m* discrete velocities in *n* dimensions. Popular examples are labeled D2Q9 and D3Q19. Higher-order lattices have been constructed [34,35] to capture some rarefied gas effects in Knudsen layer and the Knudsen paradox.

After velocity discretization, the kinetic equation (1) is reduced to a homogeneous set of linear equations in configuration space (6). To solve these equations numerically, one uses a computational grid in configuration space. Many attractive features of the original LBM method are derived for a uniform Cartesian mesh in configuration space. However, such mesh has severe practical limitations, and recent works have attempted to decouple the space mesh from the lattice structure and solve LBE equations via finite difference, finite elements and finite volume methods [21, 17]. Although some of the attractive LBM features are lost in this decoupling, other features (low degree of numerical diffusivity, etc) remain valid [21]. In practically all aspects, except for the selection method of discrete velocities, the FV LBM with decoupled space mesh and velocity lattice (see [21,17,36]) becomes similar to the FV DVM described in this paper. Another similarity is the implementation of boundary conditions using Immersed Boundary Methods (IBM) in LBM.[20]

For phase spaces with dimensions up to three (such as 1D1V or 1D2V) one can use standard mesh generation techniques and assign different axes to the corresponding configuration and velocity coordinates. Such an approach was used in [37] for Vlasov solvers on structured static mesh and in [13] for block-structured adaptive Cartesian mesh in 1D1V phase space. Figure 1, shows examples of 1D1V tree-based phase-space Cartesian meshes [38]. Such meshes are generated by subsequent divisions of unit squares (in 2D) or unit cubes (in 3D) using binary tree, quadtree, or octree data structures.

Figure 1 illustrates key differences between unsplit and split phase space grids for a 1D1V setup. The split grid assumes that binary trees are used for both configuration and velocity spaces. The split grid has planes $x$ = const. This allows simple computing of operators local in space (moments of the VDF and collisional integrals). The advection operators in $x$ and $\xi_x$ are computed independently along the $x$ and $\xi_x$ directions. On the other hand, the unsplit grid is constructed using quadtree. Computing advection operators involves a larger stencil of different direct and non-direct neighbor cells (see, e.g., [38]), which may yield a better accuracy for the advection operator, but requires special ways for the treatment of VDF moments and collisional operators. Another difference is that while a quadtree unsplit grid cell refinement proceeds along both coordinates; refinement of a split binary grid takes place independently along each coordinate (e.g., by adding new columns and rows for the 1D1V illustration in Figure 1).

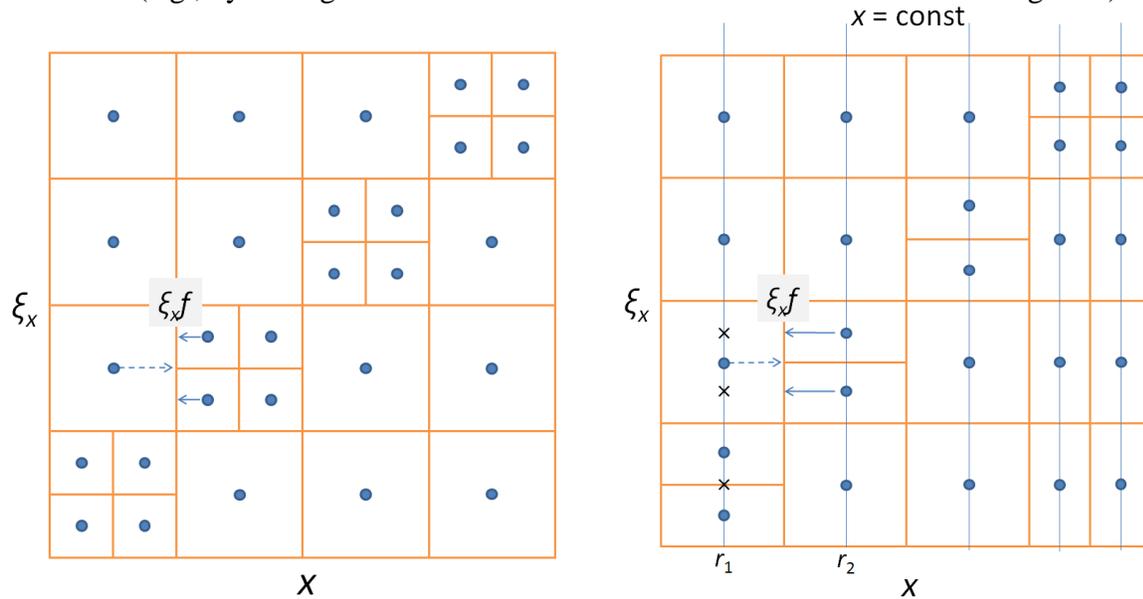

Figure 1. Examples of 1D1V phase space grids: unsplit (left, quad tree) and split binary (right, ToT). On split grid (right), all velocity cells and only last x-cell are binary divided. Arrows (solid for $\xi_x < 0$ and dashed for $\xi_x > 0$) indicate FV upwind fluxes between r-cells $r_1$ and $r_2$. Crosses indicate locations used for fine-coarse (two top crosses) and coarse-fine (bottom cross) mappings to achieve $2^{nd}$ order accuracy.

For phase spaces with dimensions larger than three special procedures need to be developed for creating velocity grids. Examples of Vlasov solvers with unsplit mesh in phase spaces of high dimensions (up to 4) and linear Boltzmann solvers with dimensions up to 5 (2D3V) have been reported [39,40]. In principle, any type of a structured or unstructured grid can be used to discretize velocity space. Hierarchical tree-based Cartesian grids have several benefits: 1) allow efficient ways to traverse and locate cells; 2) seamless grid adaptation; and 3) easy ways to develop second-order accuracy schemes for discretization of differential operators and interpolation routines. Other benefits of hierarchical Cartesian meshes are related to multi-grid and multi-level solvers. In this paper, we use tree-based adaptive Cartesian meshes for both configuration and velocity spaces.

There are important differences in implementing AMR techniques for unsplit and split phase space grids. First, for unsplit grids, mesh adaptation in configuration space triggers corresponding mesh adaptation in velocity space (for quad and octree meshes, see Figure 1, left). This could be beneficial for pure advection problems such as collisionless flows of charged particles described by the Vlasov equations. However, for collision dominated kinetics, independent grid adaptation in configuration and velocity spaces is preferable. Second, unsplit grids create additional difficulties for calculation of collisions, which involve integral operators in velocity space. Even computations of the VDF moments become cumbersome. Such computations involve summing up VDFs along $x=$const planes in $x$-$\xi_x$ space and using some interpolation techniques to find the VDF along these planes [13].

## III. TREE-OF-TREES STRUCTURE FOR PHASE SPACE

This section describes a new Tree-of-Trees framework developed for solving kinetic equations with adaptive mesh in phase space. We introduce the concept of the ToT "structure-within-a-structure", provide details of the velocity grid mapping algorithms for a treatment of advection in configuration space with velocity grids of different refinement levels, and give details about the FV discretization and time advance schemes.

### A. General Tree-of-Trees Framework

The numerical solution of partial differential equations (PDE) is based on discretization of derivatives on a computational mesh. Most PDE of mathematical physics only require a mesh in physical (configuration) space. Modern computational tools employ quad/octree meshes to simplify automatic gridding of complex geometries and adaptive mesh refinement. Kinetic equations operate in phase space of higher dimensions and require meshing for extra dimensions.

In this paper, we consider kinetic equations with phase space consisting of configuration space ($r$-space) and velocity space ($\xi$-space). Meshing this $r$-$\xi$ phase space can be done in different ways. In this work, we utilize split, unstructured $r$- and $\xi$-grids (previously illustrated in Figure *1*, right, for 1D1V case). A tree-based Cartesian grid for $r$-space was previously used as in our hybrid, kinetic-fluid solvers [24,27]. To generate an adaptive $\xi$-grid in each $r$-cell (as opposed to the previously used global, structured $\xi$-grid across all $r$-cells), we introduce a new structure-within-a-structure method. To create $\xi$-grids, we utilize the same meshing techniques as in $r$-space and generate ("grow") quad/octree $\xi$-grids in each $r$-cell, as illustrated in Figure 2. We call this concept a Tree-of-Trees (ToT).

More generally, the ToT grid with a locally varying discrete velocity grid can be considered as a six-dimensional, alternate direction tree with binary division along configuration and velocity coordinates independently ($x$ and $\xi_x$ directions for a 1D1V illustration in Figure *1*, right). Therefore, the numerical techniques developed for such binary grids can be directly used for the ToT grids of higher dimensionality. Among these techniques are higher order schemes, conservation preserving schemes, etc. However, since velocity grid is locally adapted (e.g., 1D1V $x$-$\xi_x$ space as opposed to 2D $x$-$y$ space), during operator discretization, one has to pay special attention to conservation of VDF moments (mass, momentum and energy) during the advection step. In this paper, we tackle this problem by using an accurate VDF reconstruction (mapping) technique between configuration and velocity spaces. In addition, when $r$-space adapts, $\xi$-grids need to be dynamically created and destroyed with solutions being mapped using special, local moment and property conserving fine-coarse and coarse-fine methods.

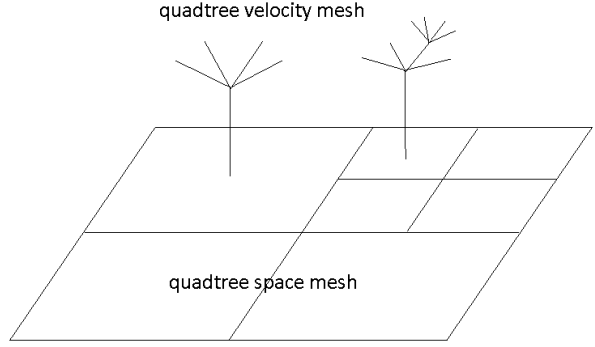

Figure 2. Tree-based $\xi$-meshes "grown" in $r$-cells, representing the concept of a Tree-of-Trees (ToT) data structure for a 2D2V case.

Therefore, the described ToT concept offers a general and flexible way to control structures within structures (such as databases, particles, etc.) on quad/octree grids, which can be used in different applications. In the remainder of this paper, we describe details of this concept which are specific to the kinetic equations for one-particle VDFs, which are controlled by advection in $r$- and $\xi$-spaces (e.g., due to external body forces) and by binary collisions in $\xi$-space. The presented ToT concept for solving kinetic equations is numerically realized within UFS methodology[24] based on the object-oriented Gfs framework.[38]

The goals of the present paper are to develop and demonstrate the ToT concept for solving the kinetic equations with adaptive mesh in phase space (AMPS) for several applications. We are aware that the initial implementation of the ToT concept carried out in this work is not numerically optimal and can be improved in several directions. There exists, e.g., an overhead related to dealing with unstructured and spatially dependent velocity grids. This overhead can be reduced by storing mapping/search data between grid adaptation events. Also, memory usage of the present ToT solver can be significantly reduced by re-writing its portions which deal with velocity space objects (grid & events). In the present paper, we give examples of speed up factors offered by the new ToT concept compared to the previously employed approaches; these factors are only approximate and often underestimated. Development of optimized ToT solvers capable of efficiently solving full 3D3V problems, and detailed study of their numerical efficiency are planned for future work.

### B. Advection in Configuration Space and Mapping between Velocity Space Grids

In our ToT solver implementation, we employ a cell-centered FV formulation where VDFs are stored on $r$- and $\xi$-cell centers. The advection in $r$-space requires calculating normal fluxes across faces of neighboring $r$-cells (with cell centers $\mathbf{r}_1$ and $\mathbf{r}_2$) for a given velocity $\boldsymbol{\xi} = (\xi_x, \xi_y, \xi_z)$, namely, $\xi_n f(t, \mathbf{r}_1, \boldsymbol{\xi})$ with $\xi_n > 0$ and $\xi_n f(t, \mathbf{r}_2, \boldsymbol{\xi})$ with $\xi_n < 0$, where $\xi_n$ is the face normal velocity. For a first-order accuracy scheme, a cell face VDF value is interpolated from its values at cell centers $\mathbf{r}_1$ and $\mathbf{r}_2$ (see also Figure 1, right). For a second-order accuracy scheme, cell-centered gradients of VDF $f$ for a given velocity $\boldsymbol{\xi}$ are calculated (with slope limiters applied to ensure solution monotonicity).

Since ξ-grids can be different in two neighboring *r*-cells with a common face (see also Figure 1 where cells $\mathbf{r}_1$ and $\mathbf{r}_2$ have different velocity grids), one needs to develop a mapping (or reconstruction[15]) technique to obtain a VDF for the same given velocity ξ in these neighboring *r*-cells. For a 2D *x-y* or 1D1V $x$-$\xi_x$ Cartesian grid, such a mapping would bring the locations where velocity and VDF need to be computed to the same *y*- or $\xi_x$-levels, correspondingly (see Figure 1, right with these locations indicated by crosses). This allows to achieve 2$^{nd}$ order accuracy (see also techniques utilized in [38] for configuration-space-only grids on unsplit grids) and our numerical experiments showed that such accuracy is adequate to accurately describe the particle kinetics under non-equilibrium conditions. (We note that for an arbitrary unstructured *x*-$\xi_x$ grid, such as a triangular grid, the velocity vector and *f* need to be reconstructed at centers of control volume faces).

While some interpolation techniques (e.g., for neighboring ξ-grids of arbitrary topology) can be used for this purpose, they can result in a loss of conservation and reduction in the scheme accuracy. In this work, we propose to use ξ-grids of the same topology (same root/building linked boxes). This way, each ξ-cell in a given *r*-cell can find a corresponding leaf, parent, or child cell in a neighboring *r*-cell. Such implementation allows exact conservation of local (in ξ-space) mass and improved conservation of local impulse and energy to at least a 2$^{nd}$-order accuracy (with errors scaling as $O(h_\xi^2)$, where $h_\xi$ is a ξ-cell size) when advecting local *f* from one *r*-cell to another. This is due to one of the following cases: 1) a leaf cell at the same level (same velocity, fine-fine type of mapping); 2) leaf cell at a lower level (fine-coarse mapping); or 3) a non-leaf cell whose cell center is at the same velocity (coarse-coarse type of mapping). In case 1 (fine-fine mapping), there is no need for a special treatment, and the VDF is directly mapped from a cell center of the corresponding ξ-cell. In case 2 (fine-coarse mapping, see Figure 3 and also Figure 1, right), the VDF can be obtained from a cell center of the coarse ξ-cell corrected by a cell-centered gradient (here, a Van Leer limited gradient is used in ξ-space to avoid introduction of non-monotone and negative VDFs). In case 3, the VDF is obtained as a result of summation over all leaf cells of the corresponding non-leaf cell (see Figure 3 and also Figure 1, right). Summation is performed over 4 cells in 2D (and 8 cells in 3D) if the difference in ξ-cell levels is 1. If this difference is 2, summation is carried out over up to 16 (e.g., 7 as in Figure 3) cells in 2D (up to 64 cells in 3D), and so on.

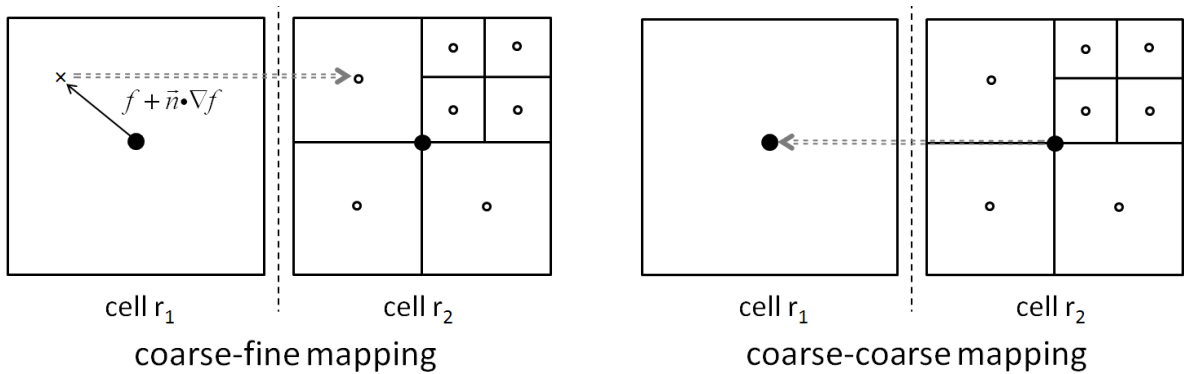

Figure 3. Examples of coarse-fine (left) and coarse-coarse (right) mapping between ξ-grids in two neighboring *r*-cells. Coarse-fine mapping involves calculation of (Van Leer limited) $\xi$-cell centered gradient and coarse-coarse mapping involves summation over all leaf $\xi$-cells (total 7 in this case) of a non-leaf $\xi$-cell in cell $r_2$.

### C. Grid Adaptation in *r*- and $\xi$-spaces

For the split *r*- and $\xi$-grids, adaptations are carried out independently (*r*-space grid adaptation results in creation of new $\xi$–space objects, and not vice versa). In the present implementation based on the object-oriented Gfs framework,[38] all events and methods available for *r*-space objects are available for $\xi$-space objects as well. In particular, grid adaptation can be easily performed in $\xi$-space based on gradients of any quantity (e.g., VDF *f* or its log). In this work, we use a simple $\xi$-space grid adaptation on gradients of VDF with a given threshold parameter. This parameter, however, is an *r*-space dependent quantity and it is scaled to the local (in *r*-space and in time) peak value of the VDF. Therefore, different $\xi$-space grid adaptation strategies can be carried out in each *r*-cell; this allows fine tuning of grid adaptation in some regions of *r*-space (such as a boundary layer near a wall, see below). We have found that for the studied problems, grid refinement variation of 3–4 levels across the *r*-space computational domain (e.g., from 4 to 7, L4-7 $\xi$-grid, or from 5 to 9, L5-9 $\xi$-grid) is adequate to obtain acceptable moment conservation and accuracy during VDF advection/mapping. Using adapted $\xi$-grids with a 3–4 refinement level range allows us to obtain large CPU time and memory gains compared to uniform $\xi$-grids at the highest level of refinement.

Grid adaptation in *r*-space requires that VDFs are dynamically created during *r*-cell refinement and destroyed during *r*-cell coarsening. During coarsening of an *r*-cell, a new $\xi$-cell object (grid & events) is created in a parent cell with a VDF computed as an average over its 4 (in 2D) children cells whose $\xi$-cell objects (grid & events) are then destroyed. During refinement of an *r*-cell, 4 (in 2D) new VDFs (or $\xi$-cell objects) are created, which are clones of the parent VDF ($\xi$-cell object). This assumes a first-order accuracy solution reconstruction algorithm. $2^{nd}$-order accuracy solution reconstruction (important for transient problems) can be readily implemented (see e.g., [38]), which requires calculation of cell-centered gradients of VDF in parent cells based on neighboring cells. Grid adaptation in *r*-space is commonly carried out on gradients of gas density, mean velocity, and/or temperature. Proper grid adaptation in *r*-space ensures that $\xi$-grids in neighboring *r*-cells remain close refinement-wise to reduce errors arising when advecting/mapping VDFs between neighboring *r*-cells.

### D. Finite Volume and Time Advance Schemes

In the current implementation of ToT kinetic solvers with the split *r*- and $\xi$-grids, an explicit time stepping scheme is used. $2^{nd}$-order accuracy in time is achieved by employing the Hancock, two-step predictor-corrector technique [41], which for a 3D3V case gives

$$f^{n+1/2} = f^n + \frac{\Delta t^n}{2} \frac{1}{h_r^3 h_\xi^3} \left( R_r[f^n] h_\xi^3 + R_\xi[f^n] h_r^3 + I[f^n, f^n] h_r^3 h_\xi^3 \right)$$
$$f^{n+1} = f^n + \frac{\Delta t^n}{h_r^3 h_\xi^3} \left( R_r[f^{n+1/2}] h_\xi^3 + R_\xi[f^{n+1/2}] h_r^3 + I[f^{n+1/2}, f^{n+1/2}] h_r^3 h_\xi^3 \right) \quad (7)$$

Here the residuals due to advection in $r$- and $\xi$-spaces are expressed as sums over faces of the control volume $r$- and $\xi$-cells:

$$R_r[f] = \sum_{r-faces} (\xi \cdot \mathbf{n}_r)_{face} f(t, \mathbf{r}_{face}, \xi) h_r^2$$

$$R_\xi[f] = \sum_{\xi-faces} (\mathbf{a} \cdot \mathbf{n}_\xi)_{face} f(t, \mathbf{r}, \xi_{face}) h_\xi^2$$

(8)

with $\mathbf{n}_r$ and $\mathbf{n}_\xi$ being the unit vectors normal to faces of the $r$- and $\xi$-cells with sizes $h_r$ and $h_\xi$, correspondingly. According to Eq.(8), the advection residuals are discretized independently on the split adaptive Cartesian $r$- and $\xi$-grids using the FV formulation. 2nd-order accuracy in space is achieved by using cell-centered gradients with (minmod or Van Leer) slope limiters to ensure monotonicity of the VDF at faces of a control volume cell (namely, at locations $\mathbf{r}_{face}$ and $\xi_{face}$, correspondingly). The values of $f$ at faces are reconstructed using the upwind scheme (see also Refs. [17,36] for the upwind schemes in LBM) both in $r$- and $\xi$-spaces with unwinding being done based on the velocity, $\xi$, and acceleration, $\mathbf{a}$, vectors, correspondingly. In order to achieve 2nd order accuracy (and thus improved conservation of the VDF moments) when reconstructing velocity and $f$ on both sides of $r$-faces (left and right fluxes) with different $\xi$-grids, we use the specially designed mapping described above. The time step $\Delta t$ is estimated from the conventional CFL criterion for both in $r$- and $\xi$-spaces taking into account the time step limitation from the collisional integral. For steady-state problems, local time stepping is used. We note that the 2nd order upwind FV cell-centered scheme with slope limiter reconstructed face values in Eq. (8) is analogous to that used in the FV LBM scheme in Ref. [17] (where a slightly different flavor of the 2nd order accuracy time-marching was used).

## IV. COLLISION INTEGRALS ON ADAPTIVE CARTESIAN VELOCITY GRID

In this Section, we describe implementation of different binary collision integrals on adaptive velocity grids. We consider two types of the collision integrals: non-linear collision integrals for elastic collisions between particles of similar types and linear collision integrals for collisions of light and heavy particles. The first type includes the model collision operators and the bi-linear Boltzmann collision integral. The second type describes elastic and inelastic collisions of light particles with heavy particles described by the Boltzmann-Lorentz integrals.

### A. Model Collision Operators

Different model collision operators can be expressed in a discrete form [8]

$$I_i = \sum_{j}^{N_v} A_{ij}(f_j - f_j^{eq}).$$

(9)

Here, $I_i = I(\xi_i)$ is the collisional integral in a cell $\xi_i$, $i$ and $j$ are the indexes of $\xi$-cells, $A_{ij}$ is a scattering matrix, and $f^{eq}$ is a local equilibrium VDF. The scattering matrix in (9) takes into account that different velocities relax to equilibrium at different rates. The matrix version of the scattering model corresponds to the multiple relaxation time model of LBM [42,43].

When the scattering matrix is reduced to a diagonal form, and a single parameter $\tau$ controls the relaxation time of all velocities, $A_{ij} = -\delta_{ij}/\tau$, where $\delta_{ij}$ is the identity matrix, one obtains a most commonly used Bhatnagar-Gross-Krook (BGK) model. The BGK model assumes $f^{eq}$ as a Maxwelian distribution. Other commonly used collision models include the Shakhov model [44] and the elliptic model [45], which differ from each other by the shape of the equilibrium VDF to achieve realistic Prandtl and Schmidt numbers not possible within the BGK model.

The significant difference of the model collision operators from the discrete Boltzmann collision integral (6) is the lack of dependency of the collision frequency on the molecular speed. Despite the fact that the model equations can often give qualitatively correct results, the degree of accuracy cannot be estimated a priory. Results for strongly nonequilibrium flows (e.g., shock waves at high Mach numbers) show substantial differences in the profiles of temperature and heat flux compared to the solutions of the full Boltzmann equation. For problems with small Mach numbers (incompressible flows) model equations give quantitatively correct solutions and their use allows significant simplification of the solution of the kinetic problem. The use of model equations of the BGK type allows calculations on a smaller number of discrete velocities by choosing a more accurate integration formula for calculation of the moments. An example of using a minimum set of discrete velocities with model collision operators is the BGK LBM.

For collisions of light particles with a heavy background lattice, $f^{eq}$ is a Maxwelian distribution with prescribed density, mean velocity and temperature. For binary collisions among particles of the same type, the Maxwellian distribution function, $f^{eq}$, should have the same local mean properties as $f$, which must be calculated by integration of $f$ over velocity space. In the latter case, the model collision operators (9) remain implicitly non-linear.

We have implemented the BGK model for our DVM solver on adaptive Cartesian velocity grids (both in 2V and 3V formulations) using special correction procedures [28]. For 2D2V situations, we introduced an additional VDF, $f_1(t,x,y,\xi_x,\xi_y)$, so that the energy contained in the full (3V) VDF is correctly taken into account]. Selected results with the BGK model for hypersonic gas flows are presented in Section V.

## B. Boltzmann Collision Integral

The procedures for calculation of the discrete collision integrals previously developed for globally defined (same $\xi$-grid across all $r$-cells) and structured Cartesian meshes must be modified for adaptive Cartesian $\xi$-grids. For numerical calculations of the discrete Boltzmann collision integral, we used a weak form of the integral [30] in the eight-dimensional space $\mathbb{R}^3 \times \mathbb{R}^3 \times 2\pi \times b_m$ with a Dirac delta function $\delta(\xi - \xi^*)$ as a test function:

$$I(\xi^*) = \frac{1}{2}\int_{\mathbb{C}} d\xi_1 \int_{\mathbb{C}} d\xi \int_0^{2\pi} d\varepsilon \int_0^{b_m} [\delta(\xi_1' - \xi^*) + \delta(\xi' - \xi^*) - \delta(\xi_1 - \xi^*) - \delta(\xi - \xi^*)]i(\xi,\xi_1)gbdb. \quad (10)$$

Here, $i(\xi,\xi_1) = f(\xi)f(\xi_1)$, $b$ and $\varepsilon$ are the impact parameters, and $b_m$ is the upper limit of the impact parameter $b$. The calculation of the collision integral takes place over a volume of velocity space $\mathbb{C} \in \mathbb{R}^3$ bounded by a sphere with the center and radius determined by the characteristic parameters of the problem.

The numerical calculation of the discrete Boltzmann collision integral poses many challenges with respect to its accuracy and efficiency. First, straightforward numerical integration is very expensive ($\sim N_v^2$). Second, only a small number of post-collision velocities on the collision sphere coincide with into ξ-cell centers even on uniform (or same level) Cartesian ξ-grids. For adaptive velocity grids, the number of such velocities tends to zero. Thus, some procedure is required to account for the velocities which do not fall exactly to the cell centers and satisfy density, momentum, and energy conservation in each collision, when taking into account the contributions of these collisions into the collision integral:

$$\int_{\mathbb{R}^3} \psi(\xi) I(f,f) d\xi = 0,$$

where $\psi(\xi) = (1, \xi, \xi^2)$ are the collision invariants. Third, the discrete collision integral must be zero for any Maxwellian distribution, $F_M$. Finally, the VDF must remain positive in all ξ-cells.

Numerous publications have been devoted to overcoming these challenges. The first challenge was addressed using Monte Carlo, quasi-random sampling [46,24], Korobov sequences [47], and importance sampling methods [48]. The second challenge was addressed using conservative projection methods. Since the majority of post-collision velocities do not coincide with cell centers of ξ-grid (especially on non-uniform ξ-grids), one has to redistribute (project) the contributions of these collisions into neighboring cells. For a uniform ξ-grid, the projection methods used in [24,46] provide good recipes for conservative calculations of the collision integral. The method of projection on two closest nodes [46] is the most economical scheme for a global (uniform) velocity mesh. In the case of a locally adaptive Cartesian ξ-grid, we extend the method of local projection into 7 closest cells first introduced in [49] for a uniform grid.

The third challenge was addressed by partition of VDF into an equilibrium part and a deviation [50,52]. All of these methods have been developed for global, uniform (structured) meshes in velocity space. In this paper, we have extended these methods for adaptive octree Cartesian ξ-grids.

## 1. Importance Sampling for Selection of Integration Nodes

Due to the high dimensionality of the collision integral (10), a Monte Carlo (MC) technique is a method of choice for its numerical integration. However, the convergence rate of conventional MC methods is quite low ($\sim 1/\sqrt{N_c}$, with $N_c$ being the number of samples), and different procedures for variance reduction are usually applied to improve the accuracy without significantly increasing the number of samples. Quasi-random Korobov sequences have demonstrated a superior performance compared to random MC sampling methods for calculations of the collision integral with global, uniform Cartesian ξ-grids. For such grids, the most expensive part - selection of collisions - is carried out only once for all *r*-cells. However, for VDFs with peculiarities in narrow regions of $\xi$-space, quasi-random selection methods lose their efficiency, even on uniform Cartesian $\xi$-grids. Moreover, these methods become inefficient for spatially dependent $\xi$-grids.

Therefore, in this paper we use the importance sampling method for selecting velocities of the colliding particles. In this approach, we perform random sampling based on a given distribution, $\Gamma(\xi)$, which is close to an integrable function and satisfies the condition $\int \Gamma(\xi)d\xi = 1$. The distribution $\Gamma(\xi)$ for Eq. (10) is defined based on the VDF, $f(\xi)$, in each $r$-cell:

$$\Gamma(\xi) = \frac{f(\xi)}{N},$$

where $N = \sum_{i=1}^{N_v} f(\xi_i)V_i$ is the particle number density, $V_i$ is the volume of a $\xi$-cell with a center $\xi_i$. To select random nodes $\xi_i$, a cumulative distribution $F(\xi_n)$ is pre-computed as

$$F(\xi_{n+1}) = \sum_{i=1}^{n} \Gamma(\xi_i)V_i$$

for $n = 1,...,N_v - 1$ and $F(\xi_1) = 0$. The function $F(\xi)$ is uniformly distributed over a unit interval (0,1). A sequence of random nodes, $\xi_l$, for $l = 1, ..., N_v$, distributed with the probability density $\Gamma(\xi)$ is determined from the solution of the equation, $F(\xi_l) = r_l$, where $r_l$ is a series of uniformly distributed random numbers. In particular, for each $r_l$ we use the bisection method to find an interval $[\xi_n, \xi_{n+1}]$ for which $F(\xi_n) < r_l < F(\xi_{n+1})$.

By choosing velocity cells $\xi_l$ based on distribution $\Gamma(\xi)$ and assuming a uniform distribution of the collision parameters ($\varepsilon, b$) (which is appropriate for isotropic scattering), we obtain the following approximation of the collision integral (10):

$$I(\xi^*) = \frac{\pi b_m}{N_c} \sum_{l=1}^{N_c} \frac{f_{1l}f_l}{\Gamma_{1l}\Gamma_l}(\Delta'_{1l} + \Delta'_l - \Delta_{1l} - \Delta_l)b_l g_l,$$

where $N_c$ is the number of collisions, and $\Delta_l$ is the approximation of a delta function on a discrete $\xi$-grid

$$\Delta(\xi - \xi_l) = \begin{cases} 1/V_l & \text{for} \quad \|\xi - \xi_l\|_\infty \leq h_\xi/2 \\ 0, \text{otherwise} \end{cases}.$$

Here $\|\xi\|_\infty = \max\{|\xi_x|, |\xi_y|, |\xi_z|\}$.

The importance sampling method was found to be particularly efficient for strongly non-equilibrium VDFs. To illustrate itsadvantages with respect to the Korobov method, we studied two simple spatially homogeneous (0D) problems using a uniform (3V) $\xi$-grid. The first problem was concerned with relaxation of two initially half-Maxwellian distributions:

$$f(\xi, t=0) = \begin{cases} F_M(N_1, u_1, T_1), \xi_x > 0 \\ F_M(N_2, u_2, T_2), \xi_x \leq 0 \end{cases} \tag{11}$$

with parameters corresponding to a shock wave (pre-shock values for $\xi_x > 0$ and port-shock values for $\xi_x \leq 0$) for given Mach number (Ma). For Ma=10 and a uniform (64×64×64) ξ-grid, the computation time using Korobov method was about 20 times greater compared to that required by the importance sampling method. The second problem was concerned with relaxation of two delta-function VDFs with non-zero values only in two cells symmetrically located with respect to the origin [12]. For this problem, Korobov method could produce no solution on a uniform (64×64×64) ξ-grid. The importance sampling method allowed solving this problem in ~20 seconds. Typical values of $N_c$ in these simulations were of the order of 20,000 for the importance sampling method and 600,000 for the Korobov method.

## *2. Multipoint Projection Method for Adaptive Cartesian Grid*

Here, we describe a generalization of the 7-point projection method [51] to non-uniform, unstructured octree ξ-grids. Figure 4 illustrates a selection procedure for such ξ-grid. Consider a post-collisional velocity ($P_\alpha$ in Figure 4) which belongs to cell $P_1$. This cell has 6 direct neighbors (leaf or non-leaf) of the same or a lower refinement level (left, right, top, bottom, front, and back). (The front and back cells are not shown in this two-dimensional figure). These 6 neighbor cells would have been selected for projection in the case of a uniform velocity mesh. In our case, however, the right and bottom neighbor cells are non-leaf cells and so have children, leaf cells. So, we select a child, leaf cell $P_2$ closest to $P_\alpha$ in the right neighbor cell, and a child, leaf cell $P_5$ closest to $P_\alpha$ in the bottom neighbor cell. The same procedure is employed for other cell-neighbor configurations.

The coefficients, $\alpha_i$, which define contributions of post-collision velocities into these cells are found for each set of velocities $\xi', \xi_1'$ from the following system of equations:

$$\sum_{i=1}^{7} \alpha_i X_i = \beta_i,$$

where $X = \{1, \xi_{xi}, \xi_{yi}, \xi_{zi}, \xi_{xi}^2 + \xi_{yi}^2 + \xi_{zi}^2\}^t$, $\beta = \{1, \xi_{x\alpha}, \xi_{y\alpha}, \xi_{z\alpha}, \xi_{x\alpha}^2 + \xi_{y\alpha}^2 + \xi_{z\alpha}^2\}^t$. In these formulas $\xi_{xi}, \xi_{yi}, \xi_{zi}, \xi_{x\alpha}, \xi_{y\alpha}, \xi_{z\alpha}$ are coordinates of the nodes $P_i$, and $P_\alpha$, correspondingly, in the local coordinate system with the center at $P_1$. To satisfy the five conservation laws (density, momentum, and energy) with seven coefficients, additional conditions are used: $\alpha_4 = \alpha_5 = \alpha_6$.

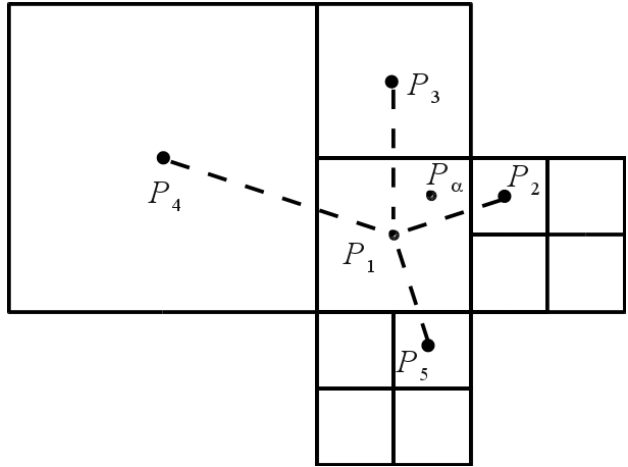

Figure 4. 2D illustration of the 7-point projection method. Front and back cell neighbors are not shown.

Using this 7-point projection method can lead to negative values of some coefficients, $\alpha_i$, which in turn can lead to negative values of the VDF. To get a positive VDF in all $\xi$-cells, a polynomial correction

$$\bar{f} = f_*(\xi)[1 + \gamma_0 + \gamma_1 \xi_x + \gamma_2 \xi_y + \gamma_3 \xi_z + \gamma_4(\xi_x^2 + \xi_y^2 + \xi_z^2)]$$

was used for the new VDF $f_*(\xi) = \max(0, f(\xi))$. The coefficients $\gamma_i$ take into account changes in density, impulses, and energy after discarding negative values of the VDF. Our test calculations showed that for the problem of homogeneous relaxation, discarding negative values of the VDF with a subsequent polynomial correction resulted in deviations that did not exceed 0.05% for the second and 0.1% for the fourth moments.

### 3. Satisfying the Equilibrium Law

The equilibrium law requires for the discrete collision integral to vanish for a Maxwellian distribution. The calculation of the collision integral (10) with contributions from direct collisions in the form $f(\xi)f(\xi_1)$ does not guarantee this requirement. This becomes particularly important for VDFs close to equilibrium. Indeed, the condition $f(\xi)f(\xi_1) = f(\xi')f(\xi_1')$ is strictly satisfied for a Maxwellian distribution. However, due to random choice of the nodes and collision parameters in the numerical evaluation of the discrete collision integral, this condition is satisfied with some error. This error comes from the fact that selection of direct collisions with velocities $(\xi, \xi_1)$ does not guarantee selection of collisions with velocities $(\xi', \xi_1')$. This reduces the values of VDF at nodes $(\xi, \xi_1)$ and increases its values at nodes $(\xi', \xi_1')$.

In the present work, we extend the approach developed in [52] for structured Cartesian grids. In this approach, the distribution function is represented as $f(\xi) = F_M(\xi) + f_d(\xi)$, where $F_M(\xi)$ is a locally Maxwellian VDF, and $f_d(\xi)$ describes a deviation from the Maxwellian. Using the property of the collision integral to vanish for any Maxwellian distribution, the weak form of the Boltzmann collision integral (10) takes the form [50]:

$$I(\xi^*) = \frac{1}{2} \int_{\mathbb{C}} \int_{\mathbb{C}} \int_0^{2\pi} \int_0^{b_m} [\delta(\xi_1' - \xi^*) + \delta(\xi' - \xi^*) - \delta(\xi_1 - \xi^*) - \delta(\xi - \xi^*)](2F_M + f_d) f_{d1} g b \, db \, d\varepsilon \, d\xi \, d\xi_1.$$

By sampling pre-collision velocities $(\xi_l, \xi_{1l})$ from the following two distributions $\Gamma(\xi) = |2F_M + f_d| / N_{2F_M + f_d}$ and $\Gamma(\xi_1) = |f_d| / N_{f_d}$, correspondingly, we approximate the collision integral as

$$I(\xi^*) = \frac{\pi b_m \rho_{2F_M + f_d} \rho_{f_d}}{M_c} \sum_{l=1}^{M_c} \text{sign}(2F_{Ml} + f_{dl}) \text{sign}(f_{d1l}) (\Delta_{1l}' + \Delta_l' - \Delta_{1l} - \Delta_l) b_l g_l,$$

where $N_{f_d} = \sum_{i=1}^{N_v} |f_{di}| V_i$, $N_{2F_M + f_d} = \sum_{i=1}^{N_v} |2F_{Mi} + f_{di}| V_i$, and the sign() terms come from the $\Gamma(\xi)$ terms. The number of collisions $N_c$ in our calculations was selected as $N_c \approx N_{2F_M + f_d} N_{f_d} g_m \Delta t / (|f_\varepsilon| V_v)$ where $|f_\varepsilon|$ is the selected minimal value of the VDF, $g_m$ is the maximal value of relative velocity, and $V_v = \pi g_m^3 / (6 N_v)$ is an average volume of a $\xi$-cell.

### C. Boltzmann-Lorentz Integral

Calculation of the Boltzmann-Lorentz collision integral (4) involves searches of all cells in velocity space which intersect a sphere with a radius $|\xi|$ centered at the origin. Each such cell contributes to the integral with a weight distributed according to the scattering law (differential collision cross section). A straightforward searching approach involves a large number of cells and is numerically expensive for large velocity grids. We therefore developed another approach, which is based on generation of $N_p$ uniformly distributed points on the $|\xi|$ sphere using the Marsaglia method [53,54]. Numerical tests showed that $N_p \sim 20$–$50$ random points on the collisional sphere were sufficient for most of the studied problems and conditions. The tests included comparisons between the results of solution of a homogeneous problem of relaxation/isotropization and of a 1D3V problem of light-particle transport which use all available points on the collisional sphere and those where a small number of randomly chosen points were used. This method resulted in significant speed up (more than 2–3 orders of magnitude) compared to the straightforward searching algorithm. We have verified that the implemented numerical algorithm conserves density and energy on octree $\xi$-grids involving cells of different sizes (refinement levels). Exact density conservation was achieved by symmetrization of scattering contributions from and into cells of different sizes. Good energy conservation was obtained by using proper grid adaptation capturing VDF spreading over the collisional sphere under the effect of collisions. For the typical cases described below, energy was conserved with less than 1–2% error. We finally note that the allowed (collisional) time step $\Delta t$ was estimated by calculating the collisional integral and then forcing the VDF to remain positive for the next time step; once the VDF starts to change (here, become isotropic) and the $\xi$-grid adapts along the collisional sphere, the time step $\Delta t$ starts to increase rapidly enabling fast time marching.

For calculations of the inelastic collision integral (5), we take into account that a particle (e.g., electron) with an initial velocity $\boldsymbol{\xi}=(\xi_x,\xi_y,\xi_z)$, after an inelastic collision, finds itself on a smaller sphere of radius $\boldsymbol{\xi}'=\sqrt{|\boldsymbol{\xi}|^2-\xi_0^2}$ distributed according to the scattering law with a weighting factor of $(\xi^2+\xi_0^2)/\xi^2$ (see Eq.(5)). Conservation of particle density in inelastic collisions was achieved by symmetrization, similar to that used in the case of elastic collisions.

## V. EXAMPLES OF SIMULATIONS AND DISCUSSION

In this section, we show examples of simulations with the newly developed ToT techniques of different physical problems in phase spaces of variable dimensionality.

### A. Charged Particle Transport in a Collisionless Sheath

In this section, we demonstrate advection in phase space in the presence of external forces, which is described by the Vlasov equation. We compare two Vlasov solvers. The first one is based on unsplit $r$-$\xi$ quadtree grid. The second solver uses the ToT method with split $r$- and $\xi$-quadtree grids.

We consider the problem of a 1D collisionless plasma sheath with particles having only one-component velocity ($\xi_x$). A space-charge sheath near a plasma boundary at a floating potential is formed to equilibrate fluxes of electrons and ions to the boundary. The sheath thickness, $L$, is of the order of the local Debye length, which is assumed to be smaller than the mean free path of charged particles. Ions enter the sheath with a Maxwellian VDF, with mean velocity equal to the ion acoustic speed, $c_s=\sqrt{T_e/M}$ (called Bohm velocity), and are absorbed at the wall. The electron VDF at the plasma-sheath boundary is assumed to be Maxwellian with zero mean velocity. The problem is characterized by two parameters: the ratio of electron to ion mass, $m/M$, and the ratio of electron to ion temperatures, $T_e/T_i$. This problem was previously analyzed using Vlasov solvers for electrons and ions coupled to a Poisson solver for self-consistent electric field [55]. Analytical solutions for the VDF of ions and electrons have been obtained for an arbitrary potential distribution in the sheath using conservation of total energy [27]:

$$f_e(x,\xi_x) = \begin{cases} C_e \exp(-\xi_x^2 + \Phi(x)), & \xi_x < \sqrt{2\Delta\Phi(x)} \\ 0, & \xi_x > \sqrt{2\Delta\Phi(x)} \end{cases}, \quad (12)$$

$$f_i(x,\xi_x) = \begin{cases} C_i \exp\left(-\frac{T_e}{T_i}\left(\sqrt{\frac{M}{m}\xi_x^2 + \Phi(x)} - 1\right)^2\right), & \xi_x > 0 \\ 0, & \xi_x < 0 \end{cases}. \quad (13)$$

Here $\xi_x = \xi_x / v_{T_e}$ is the dimensionless velocity measured in units of electron thermal velocity, $v_{T_e}=\sqrt{2T_e/m}$, $\Phi(x)$ is the dimensionless electric potential, and $\Delta\Phi(x)=\Phi(0)-\Phi(x)$. The constants $C_e$ and $C_i$ are determined by normalization conditions at the plasma-sheath boundary

($x = 0$). According to the analytical solution (12), the electron VDF is zero at $\xi_x > \sqrt{2\Delta\Phi(x)}$ because electrons are absorbed at the wall ($x = L$). The ion solution (13) corresponds to a VDF of constant amplitude and decreasing width (temperature) in the sheath.

In our simulations, for simplicity, we assume a parabolic potential profile in the sheath, $\Phi(x) = (x/2L)^2$. We first present results of computations for the ion VDF using the ToT solver and the unsplit grid solver. In the ToT solver, no grid adaptation was used in $r$-space and the $\xi$-grid adaptation was based on gradients of the VDF. The computational mesh in $r$-space and the adapted mesh in $\xi$-space (at two locations in $r$-space indicated by arrows) are shown in Figure 5. The adapted $\xi$-grid allows correct capturing the ion VDF peak position as well as its shape; see also Figure 6 (left) where the predicted ion VDFs are shown at different $x$-locations together with the analytical solution (13) at $x = L$. The amplitude of the VDF remains closely constant, and the analytical solution is reproduced with good accuracy.

For illustration and comparison, Figure 5 also shows a computational $x$-$\xi_x$ mesh and solution obtained by the unsplit Vlasov solver. The resulting solution is very close to that obtained by the ToT Vlasov solver and effectively to the analytic solution (13) (not shown for brevity).

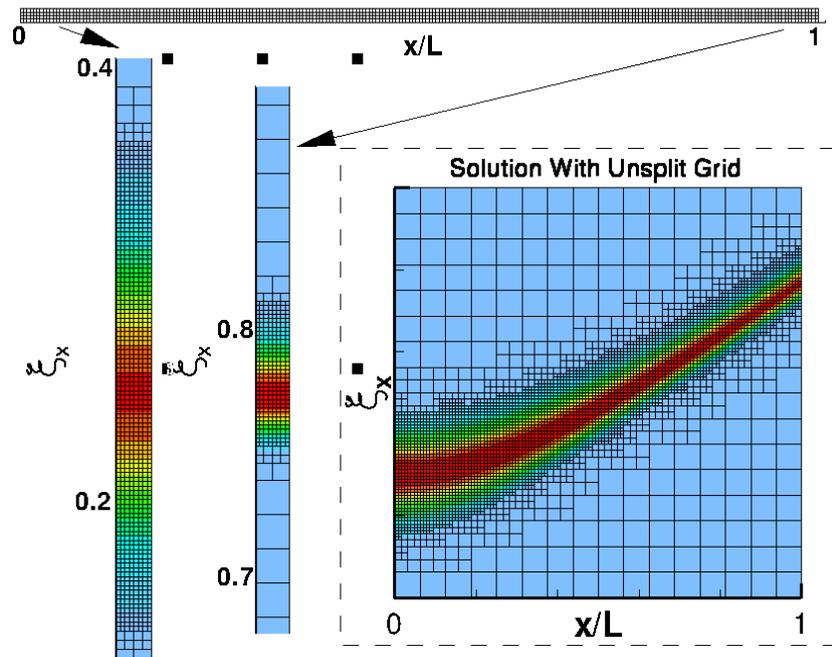

Figure 5. $r$-mesh (top) and final adapted $\xi$-mesh (left, colored by VDF) at 2 locations indicated by arrows obtained by the ToT Vlasov solver. The insert box shows the final adapted $x$-$\xi_x$ mesh and VDF obtained by the unsplit Vlasov solver.

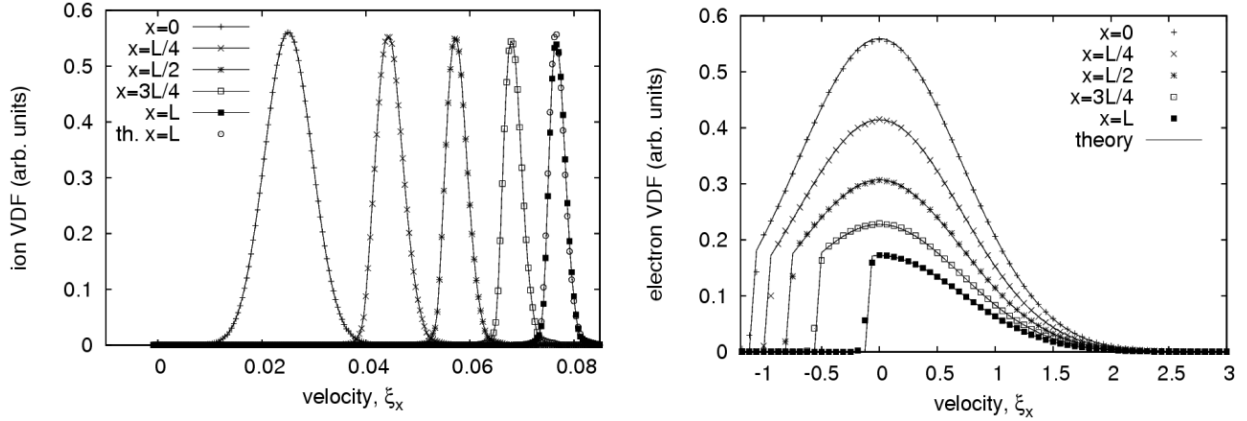

Figure 6. Ion (left, lines with symbols – numerical results, open circle symbols – analytical solution) and electron (right, symbols – numerical results, lines – analytical solutions) VDFs obtained by the ToT Vlasov solver.

Figure 6 (right) shows calculated electron VDFs at different *x*-locations, together with the corresponding analytical solution (12). Slow electrons are repelled by the electrostatic potential, fast electrons can overcome the potential well and get absorbed at the wall. The ToT Vlasov solver reproduces well the analytical solution and therefore allows for an accurate description of electron kinetics in the collisionless sheath.

### B. Rarefied Gas Dynamics

In this section, we demonstrate the ToT Boltzmann solver for rarefied gas dynamics. We consider three specific problems: spatially homogeneous relaxation (transient, 0D3V), shock wave structure (steady, 1D3V), and hypersonic flow over a square body (steady, 2D2V). The following non-dimensional units are used throughout this section. Particle velocities are normalized to the thermal velocity, $v_T = \sqrt{2RT_{ref}}$, where $T_{ref}$ is a reference (e.g., free stream) temperature. Time *t* is normalized to the collisional time, $\tau$. The Mach number for a monatomic gas is defined as $Ma = v_\infty / (v_T \sqrt{\gamma/2})$ where $v_\infty$ is the free stream velocity. The Knudsen number, *Kn*, is defined as $Kn = \lambda / L$ where *L* is a characteristic spatial scale, and $\lambda$ is the particle mean free path.

#### 1. Relaxation Problem

The implementation of discrete Boltzmann collision integral on an octree ξ-grid has been tested first for accuracy and efficiency on a problem of spatially homogeneous relaxation (0D3V setup). The initial VDF was assumed to consist of two parts: one part, for $\xi_x > 0$, corresponds to pre-shock conditions and another part, for $\xi_x < 0$, corresponds to post-shock conditions (see Eq.(11)). Computations were carried out using a single box in velocity space of size 40 (ξ-grid [-20,20]×[-20,20]×[-20,20]) for Ma = 10 and that of size 80 for Ma = 20.

Figure 7 compares the time dependence of the second and fourth moments of the VDF for Ma = 10 and 20. For these conditions, the total number of time steps to convergence was ~700–1000.

The ξ-grid was adapted on gradients of the VDF every 50[th] time step with a minimum level of 3 and a maximum level of 6 for Ma = 10 (L3-6 ξ-grid) and 7 for Ma = 20 (L3-7 ξ-grid). The results with uniform grids (L6 and L7 ξ-grids) and those with adaptive L3-6 and L3-7 ξ-grids coincide with good accuracy: the difference between the moments does not exceed 0.1%. This confirms the correctness of the implementation of the Boltzmann collision integral on octree ξ-grids which involve cells whose volumes differ by large factors (e.g., $(2^6/2^3)^3$ for L3-6 and $(2^7/2^3)^3$ for L3-7 ξ-grids).

Such a large range of grid adaptation levels yields increased efficiency of simulations. As a result of these tests, the computation time and memory required using the adaptive ξ-grid was obtained to be ~10 times smaller compared to that for a uniform ξ-grid. We note that since this test involves only one spatial (*r*) cell with complex interaction of different VDF parts (which in turn requires more refined ξ-grids), in tests involving many spatial cells with weak interaction (thus smaller ξ-grids), such as 1D and 2D shock waves, we could achieve much larger efficiencies in terms of CPU time and memory requirements (see below).

In order to demonstrate the accuracy of the developed method, we show in Figure 7 a comparison with computations using the well-validated baseline UFS-Boltzmann solver using Korobov method on a uniform Cartesian ξ-grid. It can be seen that there is a very good agreement as well, which, therefore, confirms correctness of the developed importance sampling technique combined with the multipoint projection method on adapted ξ-grids. Finally, Figure 8 shows an example of the adapted meshes and contours of the VDF at *t* = 0.32 for Ma = 10 and *t* = 0.15 for Ma = 20. It is clear that the adapted ξ-grids allow efficient capturing of VDF fine details during transient relaxation processes.

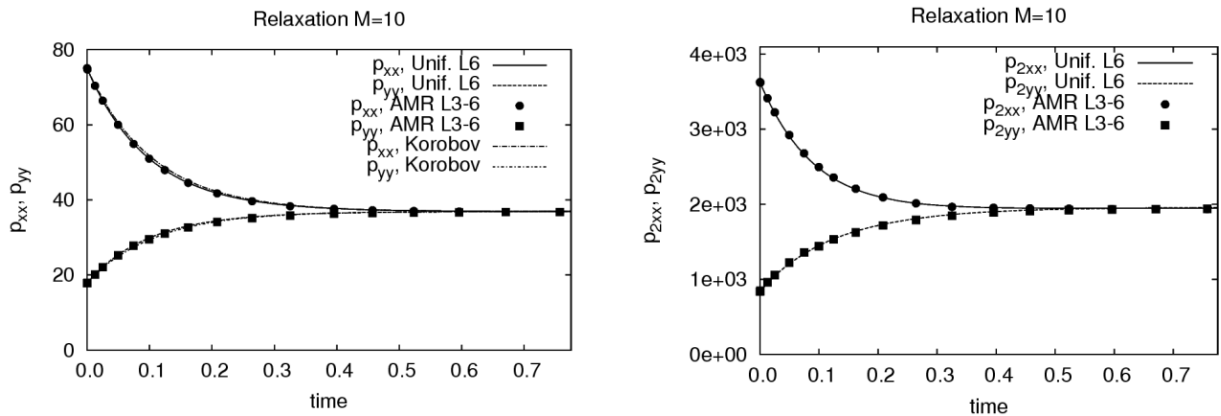

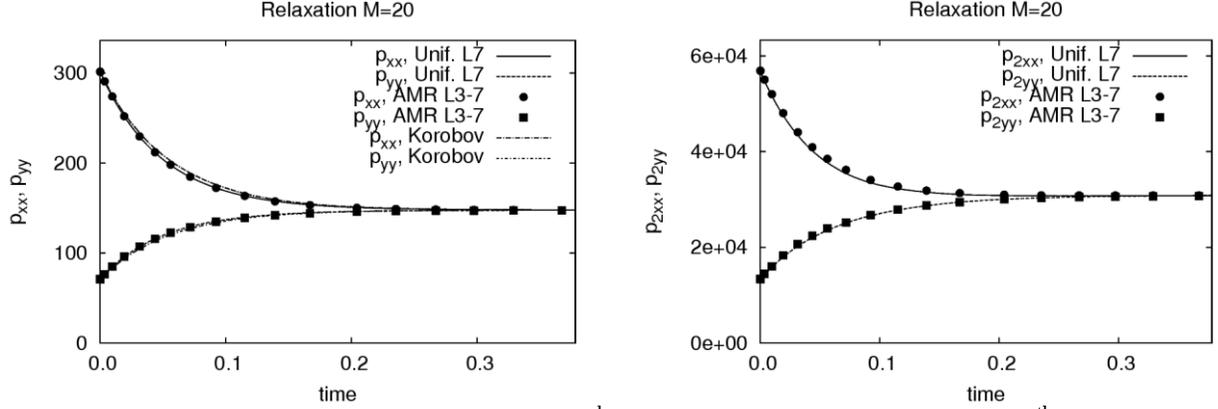

Figure 7. Time relaxation of pressure (2$^{nd}$ moment) components (left) and 4$^{th}$ moment components (right) on static (L6 and L7) and adaptive (L3-6 and L3-7) $\xi$-grids for Ma = 10 (top row) and Ma = 20 (bottom row) using the ToT Boltzmann solver on octree $\xi$-grid. Also shown are results of computations using the method of Korobov sequences on uniform $\xi$-grid.

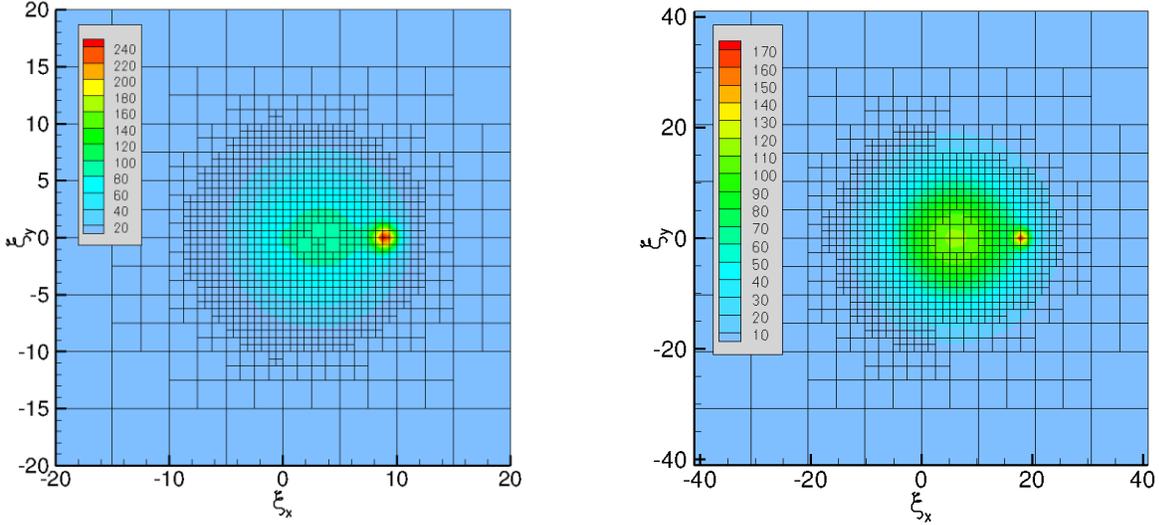

Figure 8. Adapted mesh and VDF contours for Ma = 10 (left, time 0.32) and Ma=20 (right, time 0.15) for the problem of homogenous relaxation of VDF.

## 2. Shock Wave Structure

The second test is concerned with a 1D problem of a shock wave structure in a monatomic gas with HS collisional model. In this problem, Maxwellian VDFs are specified at both ends of the computational domain in $r$-space, with parameters corresponding to jump conditions for a given Mach number. The $r$-space domain was discretized with 40 cells covering 20 mean free paths, $\lambda$, with no grid adaptation. The $\xi$-space was considered as a single box of size $[-20,20]\times[-20,20]\times[-20,20]$ (1D3V setup). Each local $\xi$-grid was adapted (on every 100$^{th}$ time step) on gradients of the VDF with a normalized threshold value of 0.2 for an L4-6 $\xi$-grid: minimum level of 4 (corresponding uniform mesh $16\times16\times16$) and maximum level of 6 (corresponding uniform mesh of size $64\times64\times64$). (We note that using a minimum level of 3 produced acceptable but less accurate results in the post-shock region). Computations were carried out until convergence was reached, which took about 2,000–3,000 time steps (dimensionless time is ~ 8–12, time step $\Delta t$ ~$5\times10^{-3}$).

The resulting profiles of density, temperature, and heat flux are shown in Figure 9. The results of computations using the baseline UFS-Boltzmann solver with the Korobov method on a uniform (structured Cartesian) $\xi$-grid are also shown for comparison. One can see that there is a very good agreement between the results providing a proof of the accuracy of the new method. Also illustrated in Figure 9 is the fact that the BGK model does not give an accurate description of the shock wave structure at high Mach numbers. Finally, we show that the number of collisions required by the new method to achieve high accuracy remains very small. Namely, about 3,000 collisions are enough outside of the transition region (where cells are in near-equilibrium) and a maximum of ~20,000 collisions are enough inside this region (with a highly non-equilibrium VDF). This yields large acceleration factors compared to the traditional methods utilizing Korobov sequences.

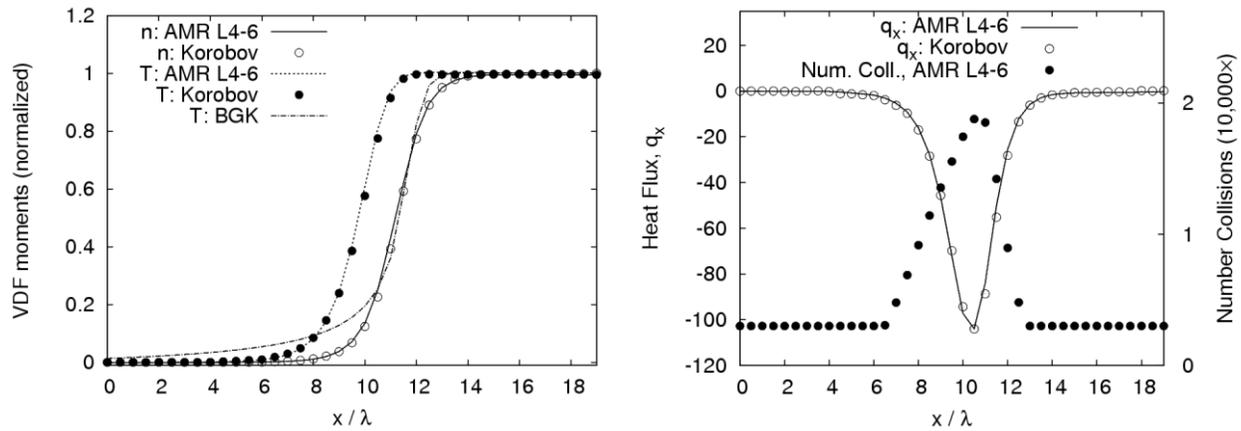

Figure 9. Comparison of the ToT Boltzmann solver on adapted $\xi$-grids with the Boltzmann solver using Korobov sequences on uniform $\xi$-grids. Left figure shows normalized gas density and temperature as functions of distance. Right figure shows heat flux and number of collisions as functions of distance. Also shown (left figure) are results of the BGK model on a L4-6 $\xi$-grid.

Adapted ξ-grids and corresponding VDFs are shown in Figure 10 at different locations inside the shock. The adapted L4-6 ξ-grid allows proper capturing of the VDF details thanks to the new method of calculating the discrete Boltzmann collision integral. Inside the shock, the VDFs are highly non-equilibrium (non-Maxwellian), while at both ends of the computational domain, the VDFs are Maxwellian with good accuracy. When a uniform ξ-grid is used for this problem, similar accuracy could only be achieved by using L6 ξ-grids. Computations with such ξ-grids in all $r$-cells become very expensive. Indeed, the CPU time and memory usage both increase by a factor of ~40 when using a uniform L6 ξ-grid. Even higher gain factors are expected for problems when the configuration domain consists of a large number of cells in near equilibrium (where small size local ξ-grids can be used) or problems with larger Mach numbers when higher resolution ξ-grids need to be used to properly resolve the VDF details.

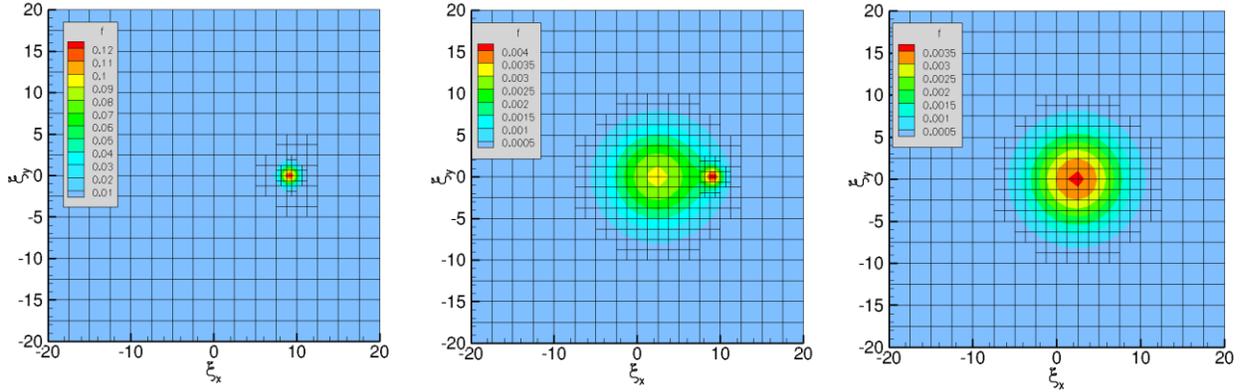

Figure 10. Adapted $\xi$-grids and VDFs, $f(\xi_x, \xi_y, \xi_z = 0)$, at different locations inside the shock: pre-shock (left), transition (middle), and post-shock (right) regions.

## 3. Hypersonic Rarefied Flows

This section presents results obtained with the developed ToT Boltzmann solver for the third test case of a rarefied hypersonic flow past a square body using the BGK model (in 2V formulation, see Ref. [28] for details). The gas is assumed to enter the computational domain through its left boundary. The boundary conditions at the body surface correspond to diffuse reflection with $T_{wall}$ = 1 (cold solid obstacle), which is equal to the free stream temperature $T_\infty$ = 1. The $r$-grid is adapted on gradients of density, mean velocity and temperature (each with its own threshold value) to ensure that the VDFs, and as a result their ξ-grids, do not differ significantly between neighboring $r$-cells. The ξ-grid adaptation in each $r$-cell is carried out on gradients of VDF with a threshold value adjusted to the local magnitude of VDF. The value of this threshold is reduced at locations near the wall so that the small amplitude reflected part of VDF (normal velocity at the wall $\xi_n < 0$) can be resolved compared to the incoming, high-speed VDF ($\xi_n > 0$). This way, the reflected portion of the VDF was well captured and proper reflection took place, consequently providing correct conditions for the formation of a bow shock around the solid obstacle.

Simulations were carried out at Kn = 0.1 for Mach numbers 10, 20, and 30. In this range of Mach numbers, the VDF shape (e.g., along the stagnation line) changes drastically. Indeed, while at lower Mach numbers (< 10) it is still numerically possible to describe the particle kinetics by a global velocity mesh, at higher Mach numbers (>10-15), the shape of the VDF changes so drastically that the use of velocity grid adaptation becomes crucial. The gas temperature contours for Ma = 10 and 30 (Ma = 20 results are omitted here for brevity) are shown in Figure 11. A bow shock region can be seen in front of the body, and the $r$-grid is well adapted to resolve this region. The flow fields along the stagnation line for Ma = 20 and 30 (Ma = 10 results are omitted here) are shown in Figure 12. We observe that the code correctly reproduces the magnitudes of flow parameter jumps across the shock wave.

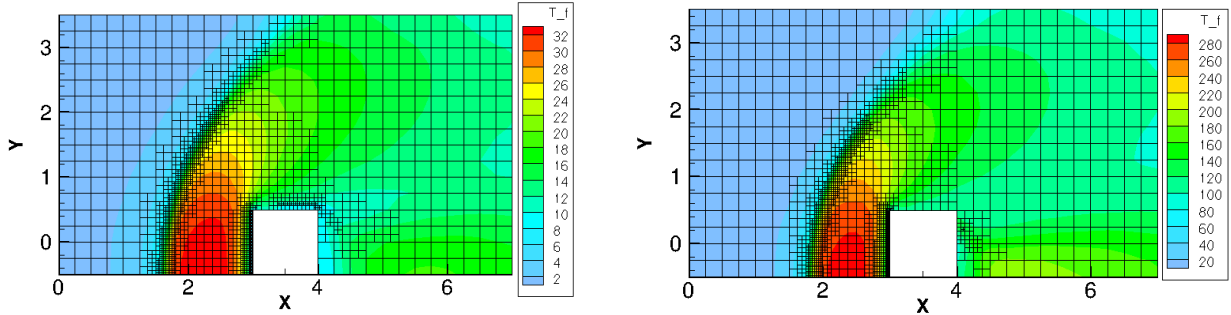

Figure 11. Spatial distributions of gas temperature and adapted $r$-grid for Ma = 10 (left) and Ma = 30 (right) for hypersonic flow over a cold ($T_{wall}$ = 1) square obstacle at Kn = 0.1.

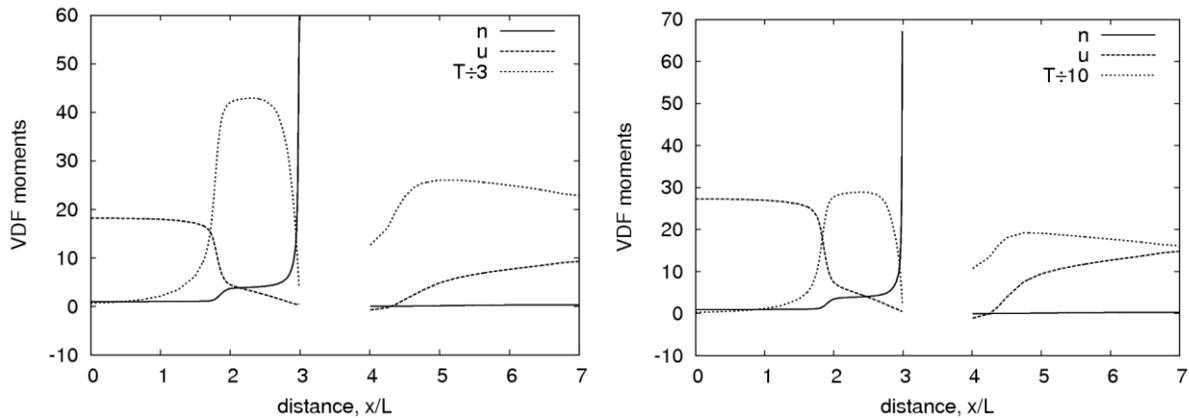

Figure 12. Gas flow density, velocity and temperature along stagnation lines for Ma = 20 (left) and Ma = 30 (right) at Kn = 0.1. (For better visibility, the plotted temperature profiles are scaled down).

The predicted VDFs at Ma = 20 are shown in Figure 13 at three locations along the stagnation line together with the corresponding adapted $\xi$-grids. One can see that the incoming VDF is very narrow in $\xi$-space and the VDF behind the shock becomes very broad. Near the wall, the VDF has another distinct feature: a jump around the $\xi_n = 0$ surface. The reflected part (at $\xi_n < 0$) is being smeared by collisions among particles due to the large value of gas density at the cold wall. The corresponding VDFs for Ma = 30 are shown in Figure 14.

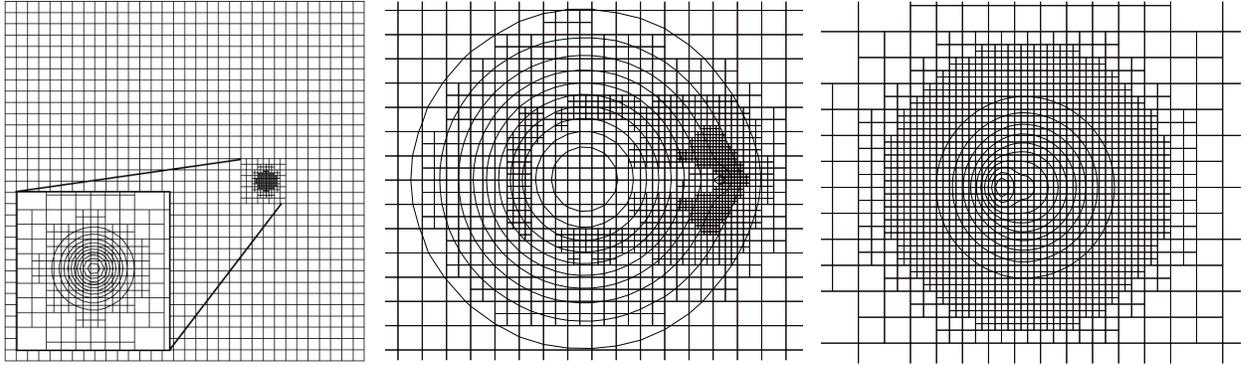

Figure 13. Adapted velocity mesh and VDF contours for Ma = 20, Kn = 0.1 at different locations: free stream (left), inside shock wave (middle) and near the wall (right).

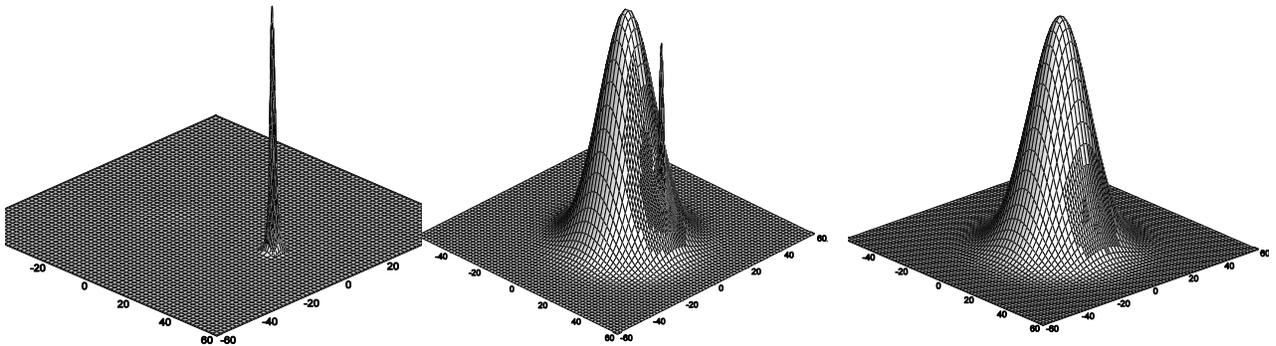

Figure 14. Adapted velocity mesh and VDFs at different locations for Ma = 30, Kn = 0.1: free stream (left), inside shock (middle), and behind shock (right).

We have carried out preliminary validation studies of the developed ToT-Boltzmann solver. Figure 15 compares the ToT solver results with DSMC results for gas macro-parameters along the stagnation line. The DSMS results were obtained with the UFS-DSMC solver [56] using the HS collision model. Despite of different collision models in the two solvers, we observed surprisingly good agreement between the gas density, mean velocity and temperature along the stagnation line, except for a region in front of the shock. The difference in the temperature profiles in this region can be attributed to the use of the BGK model in the ToT solver (see also Figure 9, left).

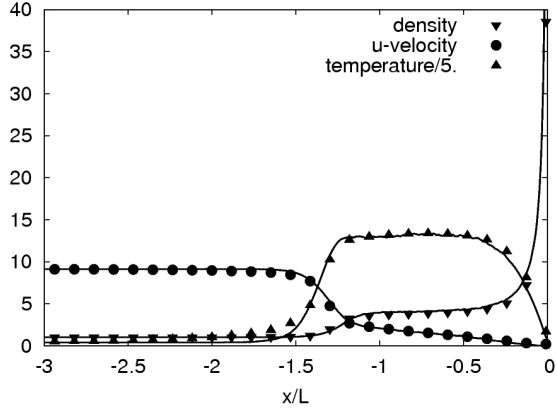

Figure 15. Comparison of ToT-Boltzmann and UFS-DSMC results for flow fields as functions of distance from the solid body along the stagnation line for Ma =10 and Kn = 0.1 conditions.

### C. Transport of Light Particles

In this section, we illustrate the AMPS technique for the linear Boltzmann equation. We first consider elastic collisions of light particles (electrons) or mass-less particles (photons) with cold heavy particles (atoms) using the collision operator in the Boltzmann-Lorentz form (4). These collisions modify direction of a particle velocity but conserve its modulus ($|\xi|$), or kinetic energy. We also consider inelastic collisions of electrons with atoms in weakly ionized plasma and emission of optical phonons in semiconductors. The numerical techniques for treatment of the Boltzmann-Lorentz collisional operator have been studied in a large number of works (see, e.g., Ref.[57] and references cited therein). Based on some prior works, it was argued[57] that finite difference schemes with Cartesian discretization are not suitable for the Boltzmann-Lorentz operator because they cannot preserve equilibrium states with a reasonable velocity mesh; the discretization errors could only be reduced by refining velocity mesh, which led to prohibitive computational cost. We show that the local velocity grid adaptation can drastically reduce computational cost compared to uniformly refined grids used in prior works and enable solving these challenging problems. Detailed studies of conservation properties of the Boltzmann-Lorentz operator on octree velocity meshes can be a subject of future efforts.

#### *1. Particle Penetration through a Thin Film*

We now present examples of the numerical solution of the linear Boltzmann equation with the Boltzmann-Lorentz collisional operator to describe radiation (photon) transport or penetration of light particles (electrons) through thin films. Consider a planar beam of photons or electrons incident normally on a thin film of thickness *L*. Details of the reflection, absorption and penetration, and the angular spectrum of the reflected and transmitted particles are determined by single scattering events [58]. The probability of single scattering is defined by the angular dependence of the collision cross section, $\sigma(\theta)$. Different approximations of $\sigma(\theta)$ used for various applications can be found in [59]. The momentum transfer cross section

$$\sigma_{tr} = 2\pi \int_0^\pi \sigma(\theta)(1-\cos\theta)\sin\theta d\theta$$

defines the value of the collision frequency $\nu = N|\xi|\sigma_{tr}$, and the particle mean free path, $\lambda = 1/(N\sigma_{tr})$. For isotropic scattering, the transport collision frequency coincides with the total collision frequency, which accounts for scattering in all possible angles:

$$\sigma_t = 2\pi \int_0^\pi \sigma(\theta) \sin\theta d\theta.$$

In our computational studies, we used isotropic scattering ($\sigma = \sigma_0 = \text{const}$) and anisotropic scattering with a simple, step-function law:

$$\sigma(\theta) = \begin{cases} \sigma_0, & \theta < \theta_{max} \\ 0, & \theta > \theta_{max} \end{cases}.$$

With decreasing $\theta_{max}$, the ratio $\sigma_t / \sigma_{tr}$ increases sharply (e.g., $\sigma_t / \sigma_{tr} \sim 200$ for $\theta_{max} = \pi/6$).

Simulations were performed in 1D3V phase space for a uniform density of scatterers ($N = \text{const}$). The incoming VDF was assumed to be a Gaussian-shaped particle beam:

$$f(x,\xi) = C_0 \exp\left\{-[(\xi_x - u_0)^2 + \xi_y^2 + \xi_z^2]/T_0\right\} \quad \text{for} \quad \xi_x > 0 \quad \text{at } x = 0,$$

where $C_0 = n_0 / (\pi T_0)^{3/2}$. A free-exit boundary condition was assumed at $x = L$:

$$f(x,\xi) = 0 \quad \text{for} \quad \xi_x < 0 \quad \text{at } x = L.$$

Most of the prior studies of this problem assumed a mono-energetic beam and so were done in $(x,\mu)$ phase space, with $\mu$ being the cosine of velocity angle. Studies are typically carried out for different collisionality degrees ranging from low (large $\lambda/L$ ratio) to high (small $\lambda/L$ ratio). Dealing with a 3V Cartesian $\xi$-grid, we model a mono-energetic beam by assuming the incoming VDFs in the limit $u_0^2/T_0 \gg 1$. In our numerical studies, we chose $n_0 = 1, u_0 = 9.12$ and $T_0 = 1$, with the ratio $u_0^2/T_0 \sim 80$ (corresponding to Ma = 10 in gas dynamics). The collisionality parameter $\lambda/L$ varied from 1/2 to 1/20. At $t = 0$, no particles are assumed to be present inside the computational domain. As the injected beam penetrates into the film, the particles are scattered according to assumed scattering law. Dynamic grid adaptation in $\xi$-space based on gradients of the VDF is carried out during our transient computations without grid adaptation in $r$-space. Below, we present converged solutions for $f(t = \infty, x, \xi)$.

Figure 16 shows the spatial distributions of the particle density, $n(x)$, at different values of $\lambda/L$ for isotropic and anisotropic scattering. One can see that for both types of scattering the density profiles flatten for larger values of $\lambda/L$. As expected, the profiles obtained for isotropic and anisotropic scattering become close for low values of $\lambda/L$ (strong scattering), and they differ most significantly at intermediate values of $\lambda/L$ (moderate scattering). The calculated profiles for different $\lambda/L$ look very similar to those obtained in Ref. [57] using a finite-element representation of the Boltzmann-Lorentz and Fokker-Planck-Lorentz operators for mono-energetic beam. Quantitative comparison is not readily possible because we used a non-monoenergitc beam in our simulations. Although our studies provide confidence in the new method, detailed validation of the Boltzmann-Lorentz operator implementation is in order.

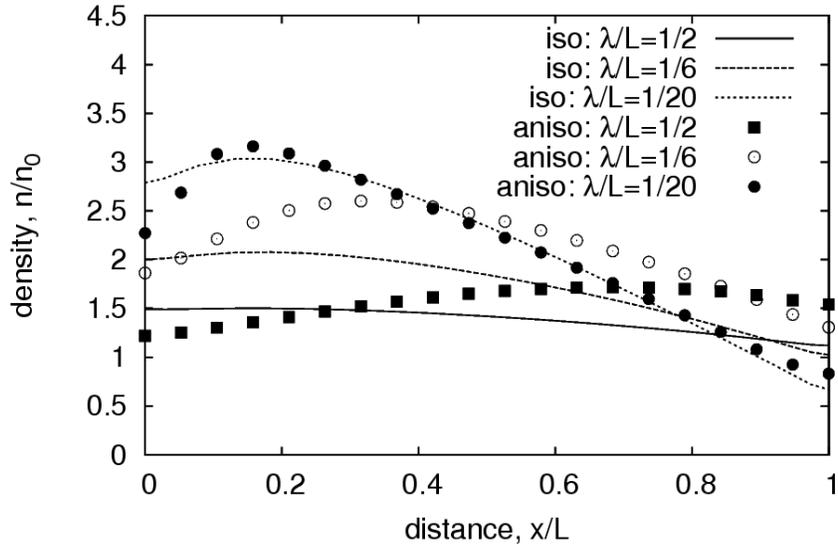

Figure 16. Light-particle density as a function of distance for different values of $\lambda/L$ for isotropic (lines) and anisotropic with $\theta_{max} = \pi/6$ (symbols) scattering.

Scattering laws are expected to have a profound impact on the angular distribution of the light particles. Figure 17 shows the VDFs near the injection point ($x = L/40$) and at the exit ($x = L$) for isotropic scattering. At the injection point, only the injected beam is present for $\xi_x > 0$ together with a scattered particles for $\xi_x < 0$. At the exit location, there is a smaller amplitude un-scattered beam together with a broad scattered wing for $\xi_x > 0$; there are no particles with $\xi_x < 0$. One can observe that the grid adaptation allows capturing the fine details of the angular distributions at different locations inside the film.

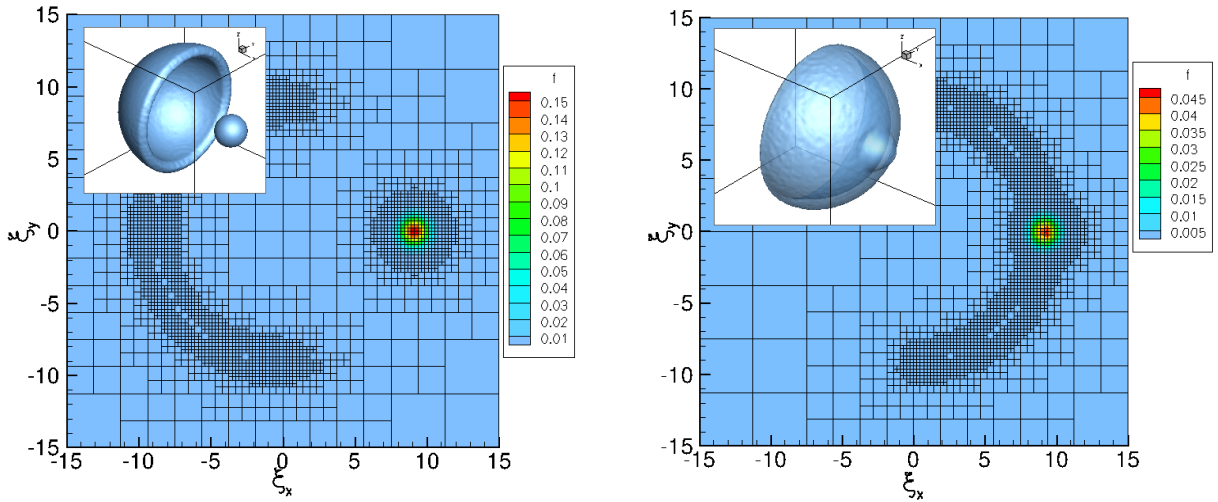

Figure 17. VDF slices at $\xi_z = 0$ and adapted $\xi$-mesh at $x = L/40$ (left) and $x=L$ (right) for isotropic scattering at $\lambda/L = 1/6$. Inserts show iso-surfaces of VDF in 3V $\xi$-space at some representative values.

Figure 18 shows the corresponding results for anisotropic scattering at $\theta_{max} = \pi/6$. In a close proximity of the injection point ($x = L/40$), the VDF for $\xi_x > 0$ consists of an injected beam and a scattered part. The scattered part is much broader compared with isotropic scattering, and fills about half of the $\xi_x > 0$ semi-sphere. This is due to the much stronger small-angle scattering at the same collisionality degree $\lambda/L$ (based on transport collisional cross-section; recall that $\sigma_t/\sigma_{tr} \sim 200$ for $\theta_{max} = \pi/6$). This leads to an increased number of small-angle scattering become important even close to the injection point. At the exit location, the VDF also differs significantly from that obtained assuming isotropic scattering. The particles fill the $\xi_x > 0$ semi-sphere almost uniformly, due to the increased role of small-angle scattering. This, in turn, results in a diffusion-like process, which corresponds to a random walk of the particles over the collision sphere.

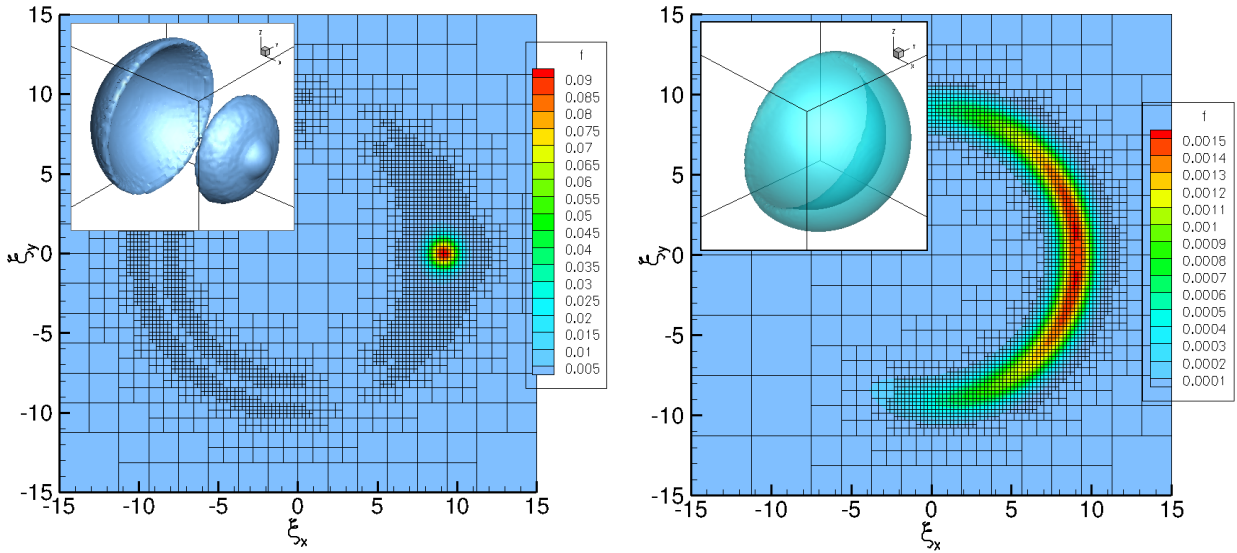

Figure 18. $\xi_z = 0$ slices of VDF and adapted $\xi$-mesh at $x = L/40$ (left) and $x = L$ (right) for anisotropic scattering and $\lambda/L = 1/6$. Inserts show iso-surfaces of VDF in 3V $\xi$-space at some representative values.

We finally demonstrate the impact of the scattering laws on the angular distribution of particles at the exit (so-called focalization effect [60]). The computed angular dependences shown in Figure 19 demonstrate very different shapes: a peaked one for isotropic scattering and a broad one for anisotropic scattering. The predicted dependences closely resemble those obtained using the Boltzmann-Lorentz and Fokker-Plank operators in the limit of a mono-energetic particle beam [60].

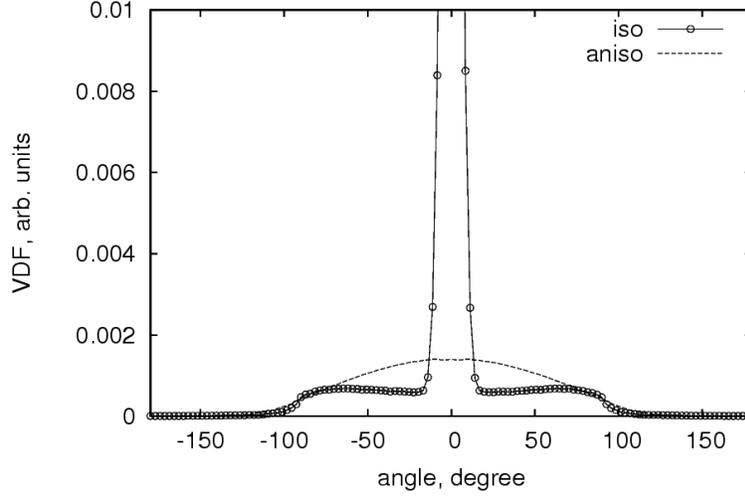

Figure 19. Comparison of angular distributions at exit location ($x = L$) for isotropic and anisotropic scattering laws for $\lambda/L = 1/6$. For clarity, VDF obtained using isotropic scattering is clipped to 0.01 (actual maximum value ~ 0.1).

To summarize this section, we have applied the AMPS technique for 1D3V problems associated with scattering of light particles from a thin film. We have calculated spatially resolved angular distributions of the particles using properly refined velocity grids and demonstrated that AMPS offers great advantages in terms of computational time and memory savings to help solving challenging problems in this field.

## 2. Electron Kinetics in Semiconductors and Gas Discharges

In this section, we demonstrate the benefits of the developed method for simulations of electron kinetics in electric fields (external force) under effect of elastic and inelastic collisions with a lattice (for semiconductors) or gas atoms (for gas discharges). In particular, we consider in detail electron streaming in semiconductors associated with formation of highly anisotropic VDFs [61]. Modeling of this phenomenon is rather difficult by the conventional methods.

To model the streaming phenomenon, we consider a simplified 0D3V problem, to study effects of electric fields and collisions on VDF formation by neglecting transport in configuration space. The electric field is assumed to be steady and directed along the $x$-axis. Both elastic and inelastic collisions are assumed to be isotropic and described by the linear operators (4) and (5), correspondingly. We assume that the characteristic time $\tau$ for elastic scattering is much larger than the time $\tau_0$ for inelastic scattering associated with emission of optical phonons. Then there are two regions in velocity space in which scattering has a completely different character; these regions are separated by a constant-energy surface, $\varepsilon(p) = \hbar\omega_0$, where $\hbar\omega_0$ is the energy of the optical phonon. If the lattice temperature $T \ll \hbar\omega_0$, then in the passive region, $\varepsilon(p) < \hbar\omega_0$, the scattering is purely elastic and is due to impurities and acoustic phonons. Meanwhile, in the active region, $\varepsilon(p) > \hbar\omega_0$, the dominant process is emission of optical phonons. In the simplest model, the boundary separating the two regions is a sphere of radius $\xi_0 = \sqrt{2\hbar\omega_0/m}$ in velocity space.

We consider a range of electric fields $E$ such that $\tau_0 \ll \tau_E \ll \tau$, where $\tau_E = m\xi_0/eE$ is the time required for electron acceleration from $\varepsilon = 0$ to $\varepsilon_0 = \hbar\omega_0$. The problem of finding VDF under these conditions is fully defined by the two ratios: $\tau_E/\tau_0$ and $\tau_E/\tau$. The streaming conditions correspond to $E/E_0 \ll 1$, where $E_0 = m\xi_0/\tau_0$ is a characteristic electric field. For our simulations, we have chosen a ratio $E/E_0 = 0.25$ (or, equivalently, $\tau_E/\tau = 4$), where the electrons are expected to form a sharply anisotropic, needle-shaped VDF.

We have carried out simulations for different values of $\nu/\nu_E$ to determine the impact of elastic collisions on the VDF shape. Figure 20 shows simulation results for three value of $\nu/\nu_E$ ratio of 0 (no elastic collisions), 1 (moderate elastic collision frequency), and 10 (large elastic collision frequency). The VDF contour plots are shown on $\xi_z = 0$ slices with white circles corresponding to the spheres $|\boldsymbol{\xi}| = \xi_0$ separating the active and passive regions. In addition, linear plots of the computed VDFs are presented in Figure 21 along $\xi_x$ (for $\xi_y = \xi_z = 0$) and along $\xi_y$ (for $\xi_x = \xi_0/2$ and $\xi_z = 0$). One can see that without elastic collisions a very narrow (needle-shaped) VDF is formed along $\xi_x$-axis in the passive region ($|\boldsymbol{\xi}| < \xi_0$), which, therefore, represents a strong streaming. In the active region, the VDF falls off sharply with a rate $\sim \tau_E/\tau$, which determines the needle width [61]. The VDF falls off as $f \sim |\xi_\perp|^{-1}$ in the transversal direction (here, $\xi_\perp = \xi_y$). At $\nu = 0$, the VDF oscillates (slightly) in time due to the transit-time resonance at frequency $\nu_E$. When moderate elastic collisions are included ($\nu/\nu_E = 1$), the needle-shaped part becomes thicker and its maximum shifts towards low velocities; the VDF outside the inelastic sphere decreases. When strong elastic collisions are considered ($\nu/\nu_E = 10$), the VDF becomes almost isotropic and its velocity dependence becomes close to $\sim |\boldsymbol{\xi}|^{-1}$, as predicted by the theory in Ref. [61]. When elastic collisions are included, the VDF converges in time without oscillations.

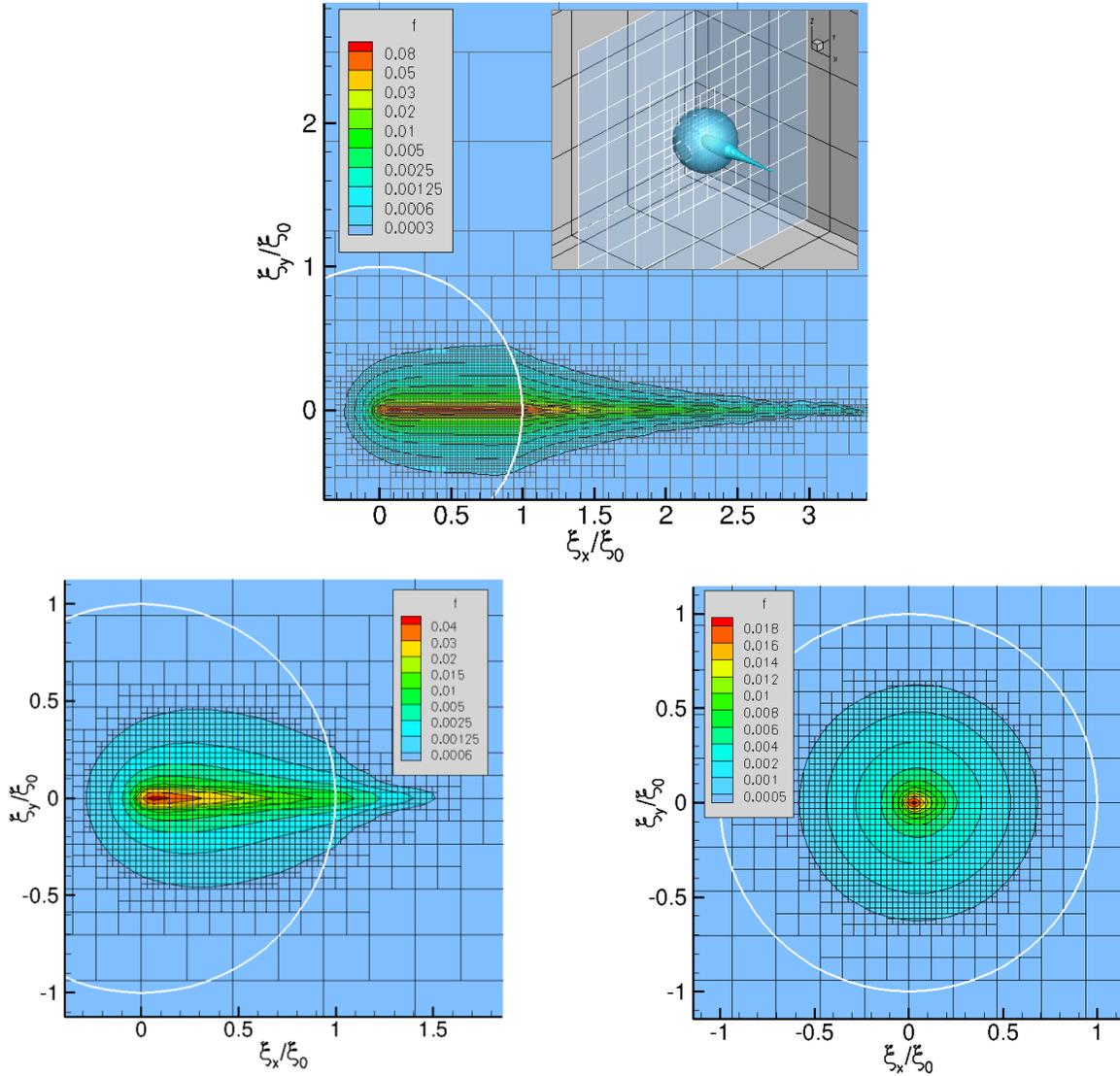

Figure 20. Computational mesh and VDF contours on $\xi_z = 0$ slices for $E/E_0 = 0.25$ and different values of $\nu/\nu_E$: 0 (top row), 1 (bottom, left), and 10 (bottom, right). White circles denote inelastic collisional sphere ($|\boldsymbol{\xi}|/\xi_0 = 1$). (Note that the contour levels are not equidistantly placed for better representation of VDF shape at higher velocities).

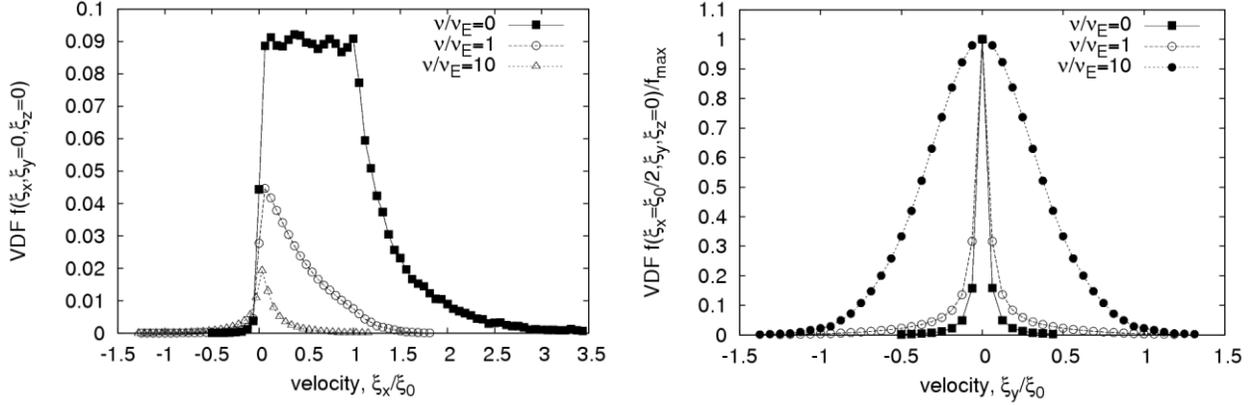

Figure 21. Linear plots of VDF for $E/E_0 = 0.25$, at different values of $\nu/\nu_E$. Left plot is for $\xi_y = \xi_z = 0$ and right plot is for $\xi_x = \xi_0/2$ and $\xi_z = 0$.

In weakly ionized plasmas of gas discharges, the corresponding collision processes are elastic collisions of electrons with neutral species, electronic excitation of atoms and molecules, and excitation of molecular vibrational levels, which are described by the collisional integrals (4) and (5). For typical fields maintaining the plasmas, conditions $E/E_0 \gg 1$ are satisfied for the vibrational excitation of molecules. The VDF formation under strong electric fields was studied in [62]. Results of our simulations for $E/E_0 = 2$ and 4 with no elastic collisions ($\nu/\nu_E = 0$) shown in Figure 22 resemble those of [62]. It is seen that the VDF consists of a small needle-shaped component at low velocities and a wide halo at larger velocities. As the ratio $E/E_0$ increases, the needle component diminishes, but it can still be traced at higher velocities even for $E/E_0 = 4$. Future work can include quantitative comparisons of the numerical computations with the theory developed in [62].

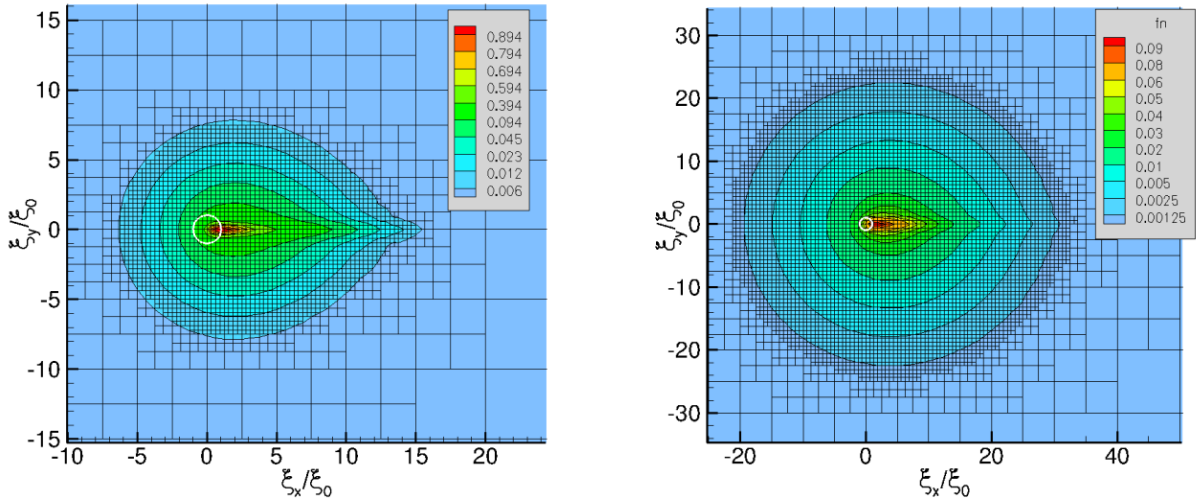

Figure 22. Computational mesh and VDF contours on $\xi_z = 0$ slices for strong electric fields: $E/E_0 = 2$ (left) and 4 (right) at $\nu/\nu_E = 0$ (no elastic collisions). White circles denote inelastic collisional sphere ($|\xi|/\xi_0 = 1$). (Note that the contour levels are not equidistantly placed for better representation of VDF shape at higher velocities).

We have, therefore, demonstrated that the developed methodology allows producing robust results for arbitrary ratios of the elastic and inelastic collision frequencies over a wide range of electric fields. It offers a noise-free alternative to Monte Carlo methods for simulations of transient processes in semiconductors, nano-structures, and gas discharges. Implementation of realistic angular dependencies of differential cross sections for anisotropic scattering and ionization processes appears to be not difficult.

## VI. CONCLUSIONS

We have introduced a novel Tree-of-Trees (ToT) concept for solving multi-dimensional kinetic equations by the discrete velocity method with adaptive mesh in phase space. An initial demonstration of this methodology has been carried out for the Boltzmann kinetic equation with non-linear and linear collision integrals for elastic and inelastic collisions. Mapping procedures have been developed to enable computations of the advection operator in configuration space on locally adapted velocity grids. Second-order accuracy in configuration and velocity spaces and in time has been achieved.

The presented FV DVM scheme for unstructured grids in configuration space was found to be analogous to the FV LBM schemes in almost all aspects (except for different ways of selecting discrete velocity sets). This connection makes the present work useful to the LBM community working on developing accurate and efficient methods for unstructured grids. Other mutual connection includes the implementation of boundary conditions using Immersed Boundary Methods (IBM).

For the first time, we have computed the bi-linear collisional operator for the discrete Boltzmann equation on adaptive velocity meshes (for the hard sphere model). Our algorithm employs several recent innovations, such as the importance sampling, multi-point projection, and variance reduction methods. Computations using well-validated baseline methods of computing the discrete Boltzmann collisional integral have shown very good accuracy and superior efficiency of the developed method. We have implemented efficient algorithms for calculating the linear Boltzmann-Lorentz collision integrals (for both elastic and inelastic collisions) of hot light particles in a Lorentz gas and the BGK collision integral on adaptive velocity meshes.

The newly developed AMPS methodology has been demonstrated for problems of hypersonic rarefied flows, light particle transport through thin films, and charged particle kinetics in plasmas and semiconductors. In particular, we considered several transient and steady-state problems in phase spaces of variable dimensionality:
1) hard sphere collisions: relaxation of particle beams, transient (0D3V);
2) advection + collisions: hypersonic flows (2D2V); supersonic shock waves (1D3V & 1D2V); transport of light particles, electrons or photons (1D3V);
3) advection + external forces: ions and electrons in collisionless plasma sheath (1D1V);
4) collisions + external force: electron kinetics in semiconductors and plasmas (0D3V).

For these problems, we have demonstrated that the AMPS technology allows minimizing the number of cells in phase space to reduce computational cost and memory usage and enable solving challenging kinetic problems. The initial implementation of this technology allows achieving speed up factors, up to almost two orders of magnitude. Higher gains are expected for larger scale and higher dimensionality problems, but extra work is required to optimize algorithms and implement parallel capabilities for solving full 3D3V problems.

We have carried an initial comparison of the ToT methodology with alternative methods using unsplit grids in phase space. The ToT method has clear advantages for kinetic solvers with binary collisions local in *r*-space. The ToT method is also beneficial for computing moments of VDF by simply traversing a $\xi$-space tree in each *r*-cell. For unsplit phase space grids, calculation of VDF moments becomes cumbersome because overlapping grid levels are not all at the same *r*-space resolution [13].

Clearly, any type of structure-in-structure representation, while favorable to the velocity-space-only operators, is less favorable for other operators, e.g., streaming in configuration space. The ToT approach is analogous to the alternate direction binary grids of higher dimensionality. This allows one to consider and extend the methods developed for configuration-space-only grids (dimension up to 3) to phase spaces of dimension up to 6. Although full flexibility of the ToT structure allows consistent mesh adaptation in phase space, additional research is required to develop synchronized mesh adaptation and VDF mapping over *r*- and $\xi$-grids. The developed methods appear to be particularly attractive for hybrid fluid-kinetic solvers because similar numerical techniques are used for both kinetic and hydrodynamic models.

## ACKNOWLEDGMENTS

This work was partially supported by the AFOSR STTR project monitored by Dr. John Schmisseur. We wish to thank Dr. Jeffrey Hittinger for useful discussions.